\documentclass[twocolumn,aps,prd,floatfix,preprintnumbers,a4paper,nofootinbib,superscriptaddress,10pt]{revtex4-1}
\usepackage{lipsum}
\usepackage{multirow}
\usepackage{import}
\usepackage[titletoc,title]{appendix}

\usepackage{graphicx}				
\usepackage{placeins}
\usepackage{amssymb}
\usepackage{amsmath}
\usepackage[dvipsnames]{xcolor}
\usepackage{subcaption}
\usepackage{verbatim}
\usepackage{rotating}
\usepackage{comment}
\definecolor{linkcolor}{rgb}{0.11,0.20,0.4}
\usepackage[colorlinks=true, linkcolor=magenta, citecolor=teal]{hyperref}
\usepackage[utf8]{inputenc}
\usepackage{txfonts}

\begin{document}

\newcommand\red[1]{{\color[rgb]{0.75,0.0,0.0} #1}}
\newcommand\green[1]{{\color[rgb]{0.0,0.60,0.08} #1}}
\newcommand\blue[1]{{\color[rgb]{0,0.20,0.65} #1}}
\newcommand\cyan[1]{{\color[HTML]{00c3ff} #1}}
\newcommand\bluey[1]{{\color[rgb]{0.11,0.20,0.4} #1}}
\newcommand\gray[1]{{\color[rgb]{0.7,0.70,0.7} #1}}
\newcommand\grey[1]{{\color[rgb]{0.7,0.70,0.7} #1}}
\newcommand\white[1]{{\color[rgb]{1,1,1} #1}}
\newcommand\darkgray[1]{{\color[rgb]{0.3,0.30,0.3} #1}}
\newcommand\orange[1]{{\color[rgb]{.86,0.24,0.08} #1}}
\newcommand\purple[1]{{\color[rgb]{0.45,0.10,0.45} #1}}
\newcommand\note[1]{\colorbox[rgb]{0.85,0.94,1}{\textcolor{black}{\textsc{\textsf{#1}}}}}
\def\gw#1{gravitational wave#1}
\def\oed#1{QA frame#1}
\def\nr#1{numerical relativity
 (NR)#1\gdef\nr{NR}}
\def\bh#1{black-hole
 (BH)#1\gdef\bh{BH}}
 \def\bbh#1{binary black hole#1
  (BBH#1)\gdef\bbh{BBH}}
\def\qa#1{quadrupole-aligned
(QA)#1\gdef\qa{QA}}
\def\pn#1{post-Newtonian
 (PN)#1\gdef\pn{PN}}
 \def\qnm#1{Quasinormal Mode
    (QNM)#1\gdef\qnm{QNM}}
   \def\eob#1{effective-one-body
      (EOB)#1\gdef\eob{EOB}}
\def\imr#1{inspiral-merger-ringdown
 (IMR)#1\gdef\imr{IMR}}
 \def\fig#1{Fig.~(\ref{#1})}
 \def\cfig#1{Fig.~\ref{#1}}
 \newcommand{\figs}[2]{Figures~(\ref{#1}-\ref{#2})}
 \def\eqn#1{Eq.~(\ref{#1})}
 \def\ceqn#1{Eq.~\ref{#1}}
 \newcommand{\Eqns}[2]{Equations~(\ref{#1})-(\ref{#2})}
 \newcommand{\eqns}[2]{Eqs.~(\ref{#1})-(\ref{#2})}
 \newcommand{\ceqns}[2]{Eqs.~\ref{#1}-\ref{#2}}
 \def\lm{{\ell m}}
 \def\tpsi{\tilde{\psi}}
\def\dcp{{\sc{PhenomDCP}}}
\def\d{{\sc{PhenomD}}}
\def\fdamp{f_1}
\def\fring{f_0}

\definecolor{brown-ish}{RGB}{181,180,20}

\newcommand{\mdh}[1]{{\color{blue}{ Mark: #1}}}
\newcommand{\sg}[1]{{\color{violet}{ Shrobana: #1}}}
\newcommand{\pk}[1]{{\color{RubineRed}{ Penny: #1}}}

\newcommand{\Cardiff}{School of Physics and Astronomy, Cardiff University, Cardiff, CF24 3AA, United Kingdom}
\newcommand{\AEI}{Max Planck Institute for Gravitational Physics (Albert Einstein Institute),
Callinstrasse 38, D-30167 Hannover, Germany}
\newcommand{\Leibniz}{Leibniz Universitaat Hannover, 30167 Hannover, Germany}
\newcommand{\BHAM}{School of Physics and Astronomy and Institute for Gravitational Wave Astronomy, University of Birmingham, Edgbaston, Birmingham, B15 2TT, United Kingdom
}

\title{First frequency-domain phenomenological model of the multipole asymmetry in gravitational-wave signals from binary-black-hole coalescence}

\author{Shrobana Ghosh}
\affiliation{\Cardiff}
\affiliation{\AEI}
\affiliation{\Leibniz}

\author{Panagiota Kolitsidou} 
\affiliation{\Cardiff}
\affiliation{\BHAM}

\author{Mark Hannam} 
\affiliation{\Cardiff}

\begin{abstract}
Gravitational-wave signals from binaries that contain spinning black holes in general include an asymmetry between the $+m$ and
$-m$ multipoles that is not included in most signal models used in LIGO-Virgo-KAGRA (LVK) analysis to date. This asymmetry manifests
itself in out-of-plane recoil of the final black hole and its inclusion in signal models is necessary both to measure this recoil, but also to
accurately measure the full spin information of each black hole. We present the first model of
the anti-symmetric contribution to the dominant co-precessing-frame signal multipole throughout inspiral, merger and ringdown. We model
the anti-symmetric contribution in the frequency domain, and take advantage of the approximations that the anti-symmetric amplitude can
be modelled as a ratio of the (already modelled) symmetric amplitude, and analytic relationships between the symmetric and 
anti-symmetric phases during the inspiral and ringdown. The model is tuned to single-spin numerical-relativity simulations up to 
mass-ratio 8 and spin magnitudes of 0.8, and has been implemented in a recent phenomenological model for use in the 
fourth LVK observing run. However, the procedure described here can be easily applied to other time- or frequency-domain models. 
\end{abstract}


\maketitle

\section{Introduction}

The LIGO-Virgo-KAGRA (LVK) collaboration has published $\sim$90 gravitational-wave (GW) 
observations~\cite{LIGOScientific:2018mvr,LIGOScientific:2020ibl,LIGOScientific:2021djp} 
since the first detection in 2015~\cite{LIGOScientific:2016aoc}. The majority of these have been from binary black holes (BBHs), 
from which we are beginning to infer the astrophysical 
distribution of black-hole masses and spins~\cite{LIGOScientific:2016vpg,LIGOScientific:2018jsj,LIGOScientific:2020kqk,KAGRA:2021duu} and references therein.
So far population inference has had to rely on limited spin information from each binary; to 
measure the magnitude and orientation of both spins we typically require louder signals than in most of those observed so 
far~\cite{Purrer:2015nkh,Khan:2020ihm}. 
We also require sufficiently accurate and physically complete theoretical waveform models. One physical effect that is
necessary to measure the full spin information is an asymmetry in the signals' multipolar structure that is not included in the standard 
full inspiral-merger-ringdown (IMR) models of \texttt{PHENOM}~\cite{Husa:2015iqa,Khan:2015jqa,Hannam:2013oca,London:2017bcn,Khan:2018fmp,Khan:2019kot,Pratten:2020fqn,Garcia-Quiros:2020qpx,Pratten:2020ceb,Thompson:2020nei,Estelles:2020osj,Estelles:2020twz} or \texttt{SEOBNR}~\cite{Taracchini:2012ig,Pan:2013rra,Taracchini:2013rva,Bohe:2016gbl,Cotesta:2018fcv,Ossokine:2020kjp,Matas:2020wab} family used in current LVK analyses.

A non-eccentric BBH is characterised by the black-hole masses $m_1$ and $m_2$, and each black hole's angular momentum, 
$\mathbf{S}_i$, which are usually represented in geometric units as the dimensionless vectors $\boldsymbol{\chi}_i = \mathbf{S}_i / m_i^2$.
The dominant effect of the spins on the GW signal is due to the spin components aligned with the binary's orbital angular momentum, 
which affect the rate of inspiral, and can therefore be measured through their effect on the signal's phase. The remaining
(in-plane) spin components have little effect on the inspiral rate. They instead induce orbital and spin precession, which lead to 
modulations in the signal's amplitude and phase~\cite{Apostolatos:1994mx}. In most cases this is a weaker contribution to the signal 
and more difficult
to measure, in turn making it difficult to measure the full spin information of the binary. Spin misalignment also leads to an asymmetry in 
the power emitted above and below the orbital plane, and can lead to large out-of-plane recoils of the final black hole~\cite{Bruegmann:2007bri}.
Most signals to date have been too weak to observe precession and recoil (with the notable exception of several analyses of the
signal GW200129\_065458 signal -- which we refer to as GW200129 in the rest of the text)~\cite{LIGOScientific:2021djp,Hannam:2021pit,Varma:2022pld}, 
but more signals with measurable spin misalignment are expected as detector sensitivities improve.

Most current generic-binary models separately consist of a model of the signal in a non-inertial frame that tracks the precession of the 
orbital plane (a ``co-precessing'' frame), and a model of the time- or frequency-dependent precession angles. If the signal is 
decomposed into spin-weighted spherical harmonics, the dominant contributions in the co-precessing frame are the 
($\ell=2, |m|=2$) multipoles. As discussed in more detail in Sec.~\ref{sec:background}, current Phenom and EOB models 
assume the symmetry $h^{\rm CP}_{22} = h^{{\rm CP}*}_{2,-2}$ in the co-precessing frame. 
Precessing binaries also include an anti-symmetric contribution. There have been 
indications for some time that neglecting this contribution could lead to measurement 
biases~\cite{Kalaghatgi:2021inv,Ramos-Buades:2020noq}, and more recently explicit 
examples of such biases have been found~\cite{Kolitsidou:2023}. 
One model that \emph{does} include the anti-symmetric contribution is 
the NR surrogate model NRSur7dq4~\cite{Varma:2019csw}, and this likely plays an important role in being able to accurately infer the primary spin in 
GW200129~\cite{Hannam:2021pit,Varma:2022pld}, demonstrating the need to include the anti-symmetric contribution in Phenom and EOB
models. 

In this paper we present a simple method to model the anti-symmetric contribution to the $(\ell=2,|m|=2)$ co-precessing-frame multipoles, 
taking into account the phenomenology of how the anti-symmetric contribution depends on the in-plane spin direction and relates to the
symmetric contribution. Note that all of the examples shown in this paper are constructed using either numerical relativity waveforms or post-Newtonian estimates and that the model introduced here is versatile in that it can be integrated into any frequency-domain approximant.

\begin{figure}[ht!]
\centering
\includegraphics[width=\linewidth]{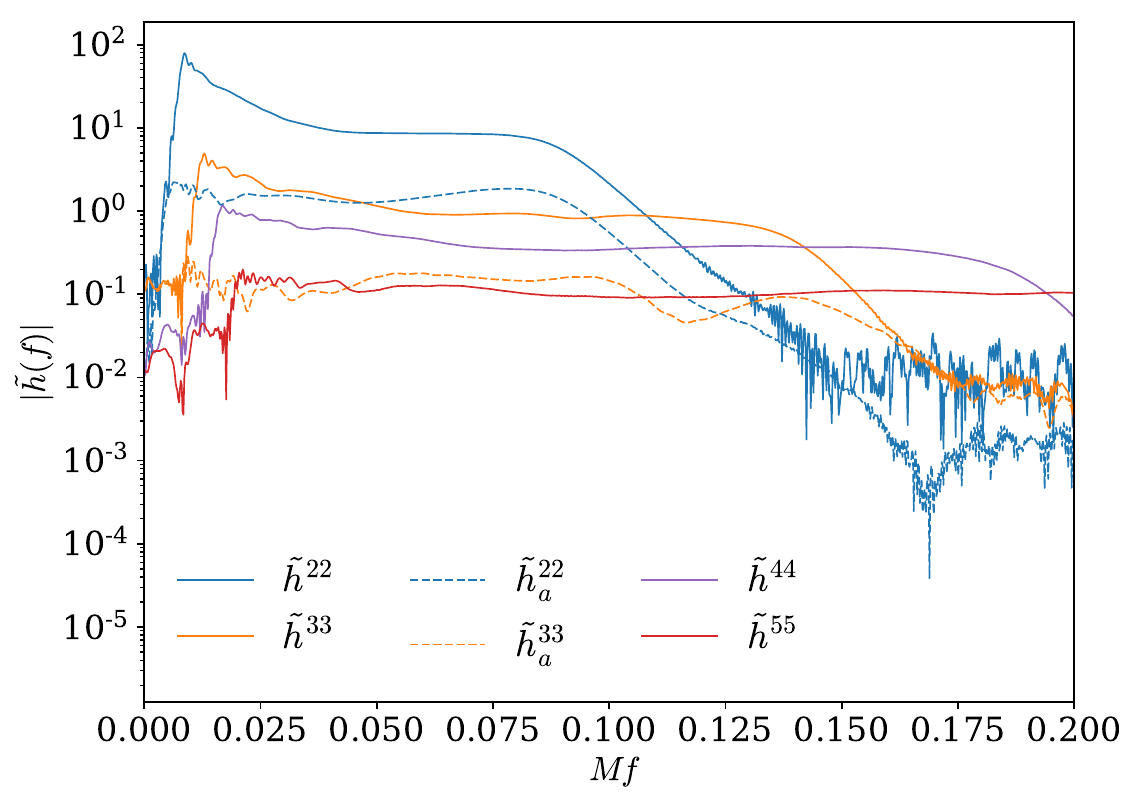}
\caption{
Frequency-domain amplitude of co-precessing-frame symmetric and anti-symmetric contributions for a $(q=2, \chi=0.8, \theta_{\rm LS} = 90^\circ)$
 configuration. The figure shows the symmetric contributions to $(2,2)$, $(3,3)$, $(4,4)$ and $(5,5)$, and the anti-symmetric contributions to
 $(2,2)$ and $(3,3)$.   
}	  
\label{fig:higher_asym}
\end{figure}

To motivate our focus on only the anti-symmetric contribution to the dominant multipoles, Fig.~\ref{fig:higher_asym} 
shows the frequency-domain amplitude of the co-precessing-frame multipoles for a signal with mass-ratio $q = m_1/m_2 = 2$, spin on the
larger black hole of $\chi = 0.8$, and spin misalignment with the orbital angular momentum of $\theta_{\rm LS} = 90^\circ$, i.e., the spin 
initially lies entirely in the orbital plane. (This is case \texttt{CF\_38} in Ref.~\cite{Hamilton:2023qkv}.)
We see that the anti-symmetric $(2,2)$ amplitude is of comparable strength to the symmetric
$(3,3)$; since the $(3,3)$ extends to higher frequencies, it will contribute more power than the anti-symmetric $(2,2)$ at high masses,
and comparable power in low-mass systems. The next-strongest anti-symmetric contribution is to the $(3,3)$, and we see that this is significantly
weaker than the symmetric $(4,4)$. This suggests that any model that includes symmetric contributions up to $\ell \leq 4$ need only include
the dominant $(2,2)$ anti-symmetric contribution. If we wish to accurately model the signal to the level of the symmetric $(5,5)$ contribution,
then we must also include the anti-symmetric $(3,3)$. Current Phenom models include symmetric multipoles up to $\ell =4$, and so we 
limit our attention to only the dominant anti-symmetric contribution. (Note that the anti-symmetric (3,3) multipole is also weaker than the 
symmetric (2,1) and (3,2) in this configuration.)

We find that we can model the $(2,2)$ anti-symmetric contribution using numerical relativity (NR) simulations
that cover only the reduced parameter space of the binary's mass ratio, the larger black hole's spin magnitude, and its misalignment
angle; to a first approximation we \emph{do not} need to sample the in-plane spin direction, which can be treated analytically. A more complete model that
includes sub-dominant in-plane-spin-direction effects, and two-spin effects, is left to future work.

In Sec.~\ref{sec:background} we explain the motivation behind our modelling approach.
We describe the preparation of the NR data that we used to calibrate our model in Sec.~\ref{section:NR_data}.
In Sec.~\ref{sec:amplitude} we construct a model of the ratio of the anti-symmetric and symmetric amplitudes, and in 
Sec.~\ref{sec:phase}, we present our method to construct the anti-symmetric phase from the symmetric phase and the
precession angle, $\alpha$. We discuss the accuracy of our prescription in Sec.~\ref{sec:matches}.

\section{Asymmetry background}
\label{sec:background}

An aligned-spin binary is invariant under reflection across the orbital plane. If we choose a coordinate system where the orbital 
plane is the $x$-$y$ plane and perform a decomposition of the gravitational-wave signal into spin-weighted spherical harmonics,
then this symmetry arises in the signal multipoles as \begin{equation}
h_{\ell,m}(t) = (-1)^\ell h^*_{\ell,-m}(t). \label{eq:symm}
\end{equation} This relationship is useful when constructing a model of the multipoles of an aligned-spin binary: we need only explicitly
model the positive-$m$ multipoles, and the negative-$m$ multipoles follow from Eq.~(\ref{eq:symm}). 

When the spins are not aligned with the orbital angular momentum, both the spins and the
orbital plane may precess~\cite{Apostolatos:1994mx,Kidder:1995zr}. In systems with mis-aligned spins Eq.~(\ref{eq:symm}) no longer holds, 
even if there is no orbital precession. The simplest illustration of this is the ``superkick'' 
configuration~\cite{Gonzalez:2007hi,Campanelli:2007cga,Bruegmann:2007bri}: here the black holes are of equal
mass and of equal spin, and the spins lie in the orbital plane but point in opposite directions. The symmetry of this system implies
that the orbital plane does not precess, and although the spins will precess \emph{in the orbital plane}, they both precess at the same rate
and so remain oppositely directed to each other, so that the vector sum of the two spins is zero at all times. Although the direction of the 
orbital plane remains fixed, emission of linear momentum perpendicular to the orbital plane causes the entire system to move up and
down. This linear momentum emission and ``bobbing''~\cite{Keppel:2009tc} manifests itself in the gravitational-wave signal as an 
asymmetry in the positive- and negative-$m$ multipoles, i.e., a violation of Eq.~(\ref{eq:symm})~\cite{Bruegmann:2007bri}.

The symmetry of Eq.~(\ref{eq:symm}) will remain broken regardless of any rotations performed on the multipoles~\cite{Boyle:2014ioa}. 
This point becomes important when constructing signal models, where we regularly make use of a ``co-precessing frame''. In this
frame the signal during the inspiral can be approximated as that of a non-precessing binary~\cite{Schmidt:2012tmg} and so
many current waveform models are split into a
model for aligned-spin binaries and a model for the precession dynamics, and the precession dynamics are then used to ``twist up'' 
a non-precessing-binary waveform to produce the complete precessing-binary waveform~\cite{Hannam:2013oca,Khan:2018fmp,Khan:2019kot,Pratten:2020ceb,Estelles:2021gvs,Pan:2013rra,Taracchini:2013rva,Ossokine:2020kjp,Ramos-Buades:2023ehm}. However,
since the non-precessing-binary waveform respects the symmetry in Eq.~(\ref{eq:symm}), the model cannot reproduce the asymmetry
that should be present in the true precessing-binary signal. 

Several studies have considered the impact of neglecting these multipole asymmetries. Ref.~\cite{Ramos-Buades:2020noq} 
compares the multipoles from precessing-binary waveforms with those from nominally equivalent non-precessing binaries, to test
a number of assumptions that go into the construction of many commonly used waveform models, including neglecting the 
multipole asymmetry.  Ref.~\cite{Kalaghatgi:2021inv} 
compares NR waveforms from configurations with different in-plane spin directions and magnitudes, and argues that neglecting
the multipole asymmetry may lead to parameter biases even at moderate SNRs, and that including the multipole asymmetry 
will be necessary to clearly measure in-plane spins and identify precessing systems. Finally, Ref.~\cite{Kolitsidou:2023} uses the 
surrogate model NRSur7dq4 to identify the level of bias in binary 
measurement examples, and confirms that neglecting the multipole asymmetry leads to biases in measuring in-plane spins
(but the masses and effective aligned spin $\chi_{\rm eff}$ are less affected). They also confirm the importance of the multipole asymmetry 
in precession measurements, showing that it had a significant impact on measurement of the properties of the signal 
GW200129~\cite{Hannam:2021pit,Varma:2022pld}. 

In the next section we will summarise the leading-order PN contribution to the asymmetry, which provides some insight into the
phenomenology of the multipole asymmetry, and also motivate our modelling procedure. Although the multipole asymmetry has
been known for some time, and indeed is included in the standard PN expressions that we use here, and is also
discussed in detail in Ref.~\cite{Boyle:2014ioa}, we are not aware of any prior treatment that discusses the amplitude and
phasing of the anti-symmetric (2,2) contribution in relation to the symmetric contribution, or notes the simple dependence
of the relative phase between different in-plane spin directions, which is a key feature of the asymmetry that we exploit in 
constructing our model.

\subsection{Inspiral}

To gain insight into the phenomenology of the multipole asymmetry during the inspiral, we consider the leading-order post-Newtonian contributions
to a binary where only one black hole is spinning and the spin lies entirely in the orbital plane.
The binary consists of two black holes with masses $m_1$ and $m_2$ and the dimensionless spin on the primary is 
$\chi = S_1 / m_1^2$, where $S_1$ is the magnitude of the black hole's angular momentum. We use the 
post-Newtonian expressions from Ref.~\cite{Arun:2008kb}, where in this single-spin case the symmetric and anti-symmetric
spin contributions are $\chi_s = \chi_a = \chi/2$. 
The in-plane spin components incline the total angular
momentum $\mathbf{J}$ with respect to the normal to the orbital plane (and direction of the Newtonian orbital angular momentum, 
$\mathbf{L}$) by an angle $\iota$, and the azimuthal precession angle of $\mathbf{L}$ around $\mathbf{J}$ is $\alpha$; this is also the 
azimuthal angle of the total in-plane spin. As such, if we choose the instantaneous orbital plane to coincide with the $x$-$y$ plane, then the 
entirely in-plane spin can be written as $\boldsymbol{\chi} = \chi (\cos(\alpha),\sin(\alpha),0)$. 

We start with the multipole $h_{22}$ as given in Eq.~(B1) in Ref.~\cite{Arun:2008kb}. Requiring symmetry due to exchanging
black holes (see the discussion prior to Eq.~(4.15) in the same paper), leads to the relation 
$h_{\ell m}(\Phi) = (-1)^{\ell + m} h_{\ell -m}^* (\Phi + \pi)$, where $\Phi$ is the orbital phase. 
We can enter the instantaneous orbital plane by setting $\iota = \alpha = 0$, and from our choice of spin we can then substitute 
$\chi_{ax} = \chi_{sx} = \chi \cos(\alpha)/2$ and
$\chi_{ay} = \chi_{sy} = \chi \sin(\alpha)/2$; see sec. VI.B of Ref.~\cite{Hamilton:2021pkf} for details.
We then have an approximation to the symmetric and anti-symmetric 
contributions to the signal in the co-precessing frame, $h_{22}^{\rm{CP},s} = ( h_{22}^{\rm CP} + h_{2-2}^{*\rm CP})/2$ and $h_{22}^{\rm{CP},a} = ( h_{22}^{\rm CP} - h_{2-2}^{*\rm CP})/2$, up to $O(v^4)$, 
\begin{eqnarray}
h_{22}^{\rm{CP},s} & = & A  \left( 1 + \frac{  (55 \eta - 107) v^2}{42} \right) e^{-2i \Phi}, \label{eq:PNsym} \\
h_{22}^{\rm{CP},a} & = & A \frac{ (1 + \delta) \chi \, v^2}{4} e^{-i (\Phi + \alpha)} \label{eq:PNasym},
\end{eqnarray}
 where the overall amplitude is $A =  \sqrt{ 64\pi/5}  M \eta v^2/ D_L$, $M$ is the total mass,
$\eta = m_1 m_2 / M^2$ is the symmetric mass ratio, $\delta = (m_1 - m_2)/M$, $v$ is the relative speed of the two black holes, 
and $D_L$ is the luminosity distance to the source. Note that in Ref.~\cite{Arun:2008kb} the symbol $\Phi$ denotes the 
orbital phase in the instantaneous orbital plane, but here we use it to denote the total orbital phase that enters in the waveform. 

We may immediately note several important features from Eqs.~(\ref{eq:PNsym}) and (\ref{eq:PNasym}). The spin does not enter
the amplitude of the symmetric contribution at this order. The anti-symmetric contribution enters at $O(v^2)$ lower than the symmetric
contribution. We see that we may also consider the anti-symmetric amplitude $|h_{22}^{\rm{CP},a}|$ as a simple rescaling of the 
symmetric amplitude, $|h_{22}^{\rm{CP},s}|$. 

The in-plane spin direction, $\alpha$, does not enter into the amplitude of the anti-symmetric contribution, but does modify the phase. 
The physical interpretation is that the phase of the anti-symmetric contribution depends on the direction of the in-plane spin relative
to the separation vector of the two black holes. This will vary with the orbital phase, $\Phi$, but also the (slower) precession rotation of 
the spin, given by $\alpha$. This is also consistent with the observation in studies of out-of-plane recoil, that the recoil amplitude 
depends sinusoidally on the initial direction of the in-plane spin~\cite{Bruegmann:2007bri}. 

Finally, we note a key observation for the 
model that we will produce: if we modify the initial in-plane spin direction by $\Delta \alpha$, this will induce a simple overall phase shift
in the anti-symmetric contribution, $h_{22}^{\rm{CP},a}$. This suggests that, given a set of single-spin numerical-relativity (NR) waveforms 
that cover the parameter space of mass ratio, aligned-spin magnitude and in-plane spin magnitude, we will have enough information to build
a model of the anti-symmetric contribution to single-spin waveforms \emph{without the need to also sample multiple initial in-plane spin directions}.
We have just such a set of waveforms to hand, as used to construct the first NR-tuned full inpiral-merger-ringdown model of 
(the symmetric contribution to) precessing-binary waveforms~\cite{Hamilton:2021pkf}, and discussed in 
Ref.~\cite{Hamilton:2023qkv}.

\subsection{Merger and ringdown}
\label{sec:imr}

Before proceeding to construct a model, we consider the phenomenology of the anti-symmetric contribution 
through merger and ringdown, and inspect which inspiral features hold for the entire waveform. 

\begin{figure}[ht!]
\includegraphics[width=\linewidth]{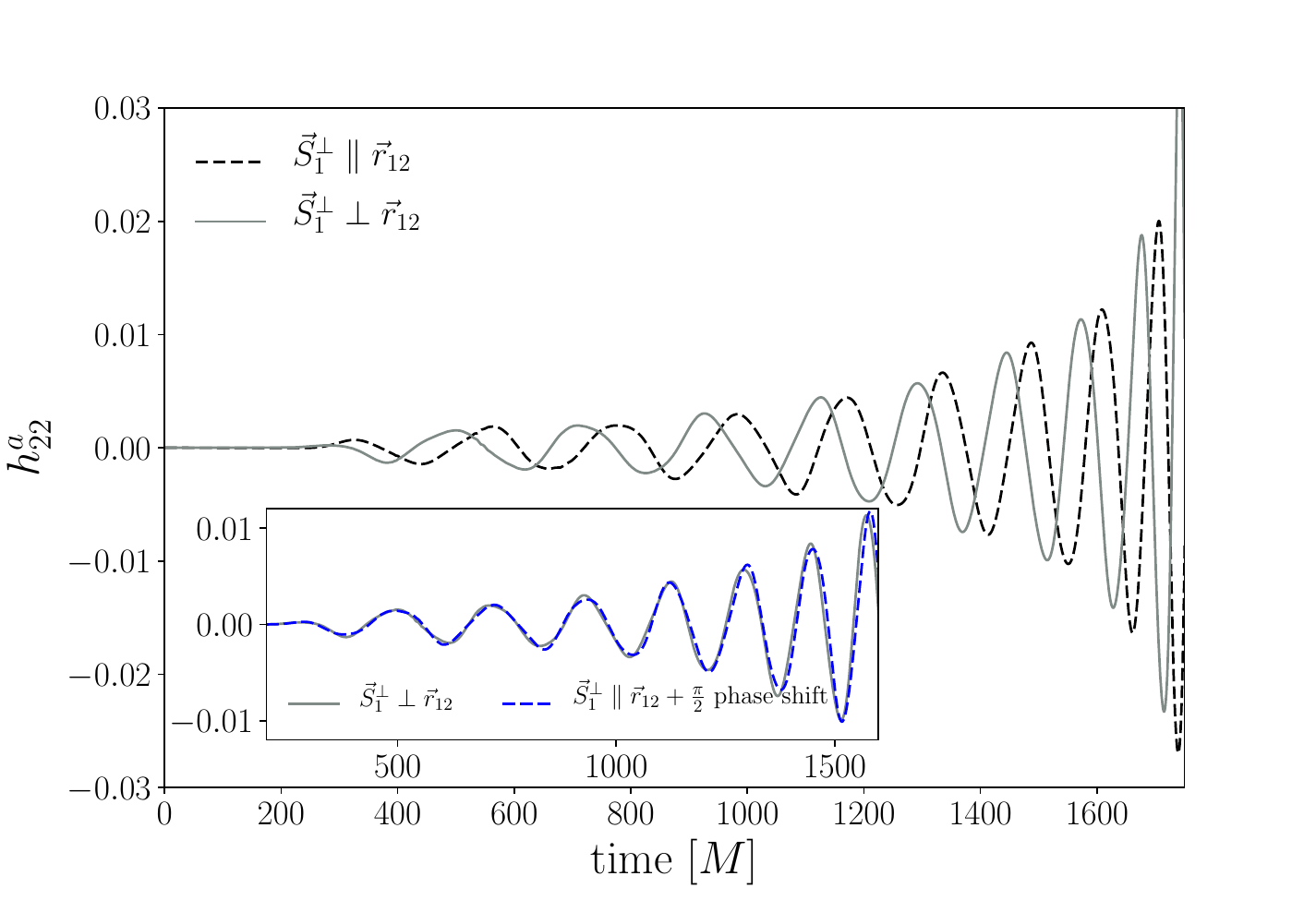}
\caption{Anti-symmetric waveform constructed from NR waveforms for two superkick configurations in the time-domain, $\vec{S}^{\parallel}_{1} \perp \vec{r}_{12}$ in dashed black and $\vec{S}^{\perp}_{1} \perp \vec{r}_{12}$  in solid grey, shows a constant phase offset of $\pi/2$. \textit{Inset}: Dashed blue line shows that the anti-symmetric waveform for $\vec{S}^{\perp}_{1} \perp \vec{r}_{12}$ can be constructed by just applying a $\pi/2$ phase shift to the $\vec{S}^{\perp}_{1} \parallel \vec{r}_{12}$ waveform even in the strong-field regime close to merger $(t_{\rm{merger}}=1784 M)$.}
\label{fig:spinrotation}
\end{figure}

One of the main features of the anti-symmetric contribution that we see in the leading-order inspiral single-spin 
expressions (\ref{eq:PNsym}) and (\ref{eq:PNasym}) is that an in-plane rotation of the spin by an angle $\Delta \alpha$ 
results in a corresponding shift in the anti-symmetric (2,2) phase by $\Delta \alpha$. This is evident from 
Fig.~\ref{fig:spinrotation}; the two configurations considered here correspond to the superkick configuration described earlier. 
$\vec{S}^{\perp}_{1}$ denotes the initial in-plane spin vector of the primary and $\vec{r}_{12}$ is the initial separation vector pointing from 
the primary to the secondary.  It is clear that the asymmetry phase for a configuration with $\vec{S}^{\perp}_{1} \perp \vec{r}_{12}$ can be 
easily produced by applying a phase shift of $\pi/2$ to the anti-symmetric waveform of a configuration with 
$\vec{S}^{\perp}_{1} \parallel \vec{r}_{12}$. We note that the simple phase relationship does not appear to hold as well through merger and ringdown, but the  deviation is small enough that this could be due to numerical error, and requires more detailed study in future.

\begin{figure}[ht!]
\includegraphics[width=\linewidth]{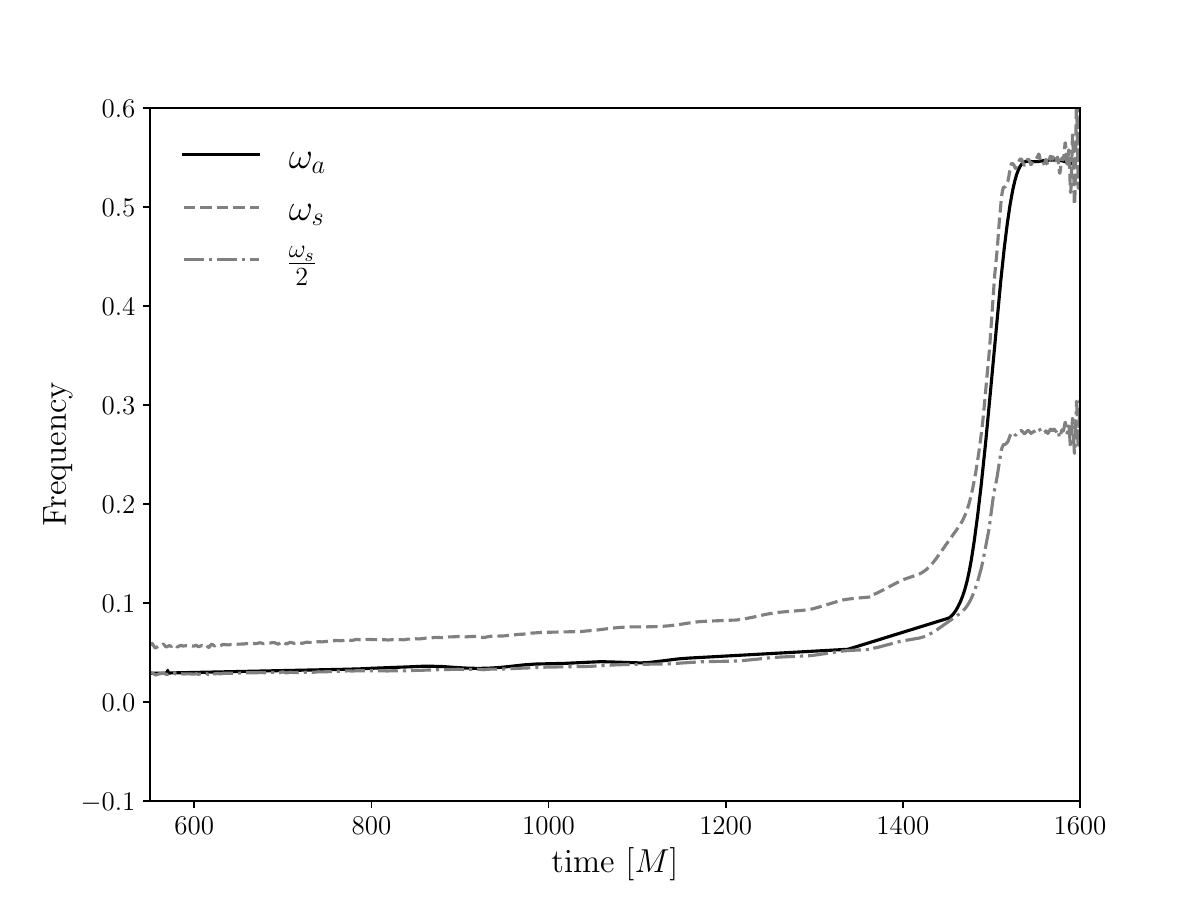}
\caption{The symmetric and anti-symmetric frequencies obtained from NR data of a $(q=2, \chi=0.7, \theta_{\rm LS} = 90^\circ)$ configuration. 
During inspiral the anti-symmetric frequency, $\omega_a$, is about half of the symmetric frequency, $\omega_s$ (or nearly equal to the orbital frequency), and close to merger quickly catches up with the symmetric frequency. As expected from perturbation theory,
$\omega_s = \omega_a$ during ringdown.}
\label{fig:asymfrequency}
\end{figure}

A second key feature of the anti-symmetric contribution is that its frequency is roughly half that of the symmetric 
contribution (plus a small correction from the spin-precession rate $\dot{\alpha}$).  Fig.~\ref{fig:asymfrequency} shows the time-domain 
GW frequency of the symmetric and anti-symmetric contributions for a configuration where only the larger black hole is spinning, with the spin ($\chi=0.7$) entirely in-plane. 
We see that during the inspiral the anti-symmetric frequency is approximately half that of the symmetric, as we expect. 

During ringdown the two frequencies are equal. This is consistent with our expectations from perturbation theory. 
In the ringdown, where perturbation theory results are applicable, the $(\ell=2,m=\pm2)$ multipoles will 
have the same (constant) frequency, and will decay exponentially at the same rate. We therefore expect both the 
symmetric and anti-symmetric combinations of $h_{22}$ and $h_{2,-2}$ to have the same frequency, and for the 
ratio of the symmetric and anti-symmetric amplitudes to be constant throughout the ringdown. 

\begin{figure}[ht!]
\includegraphics[width=\linewidth]{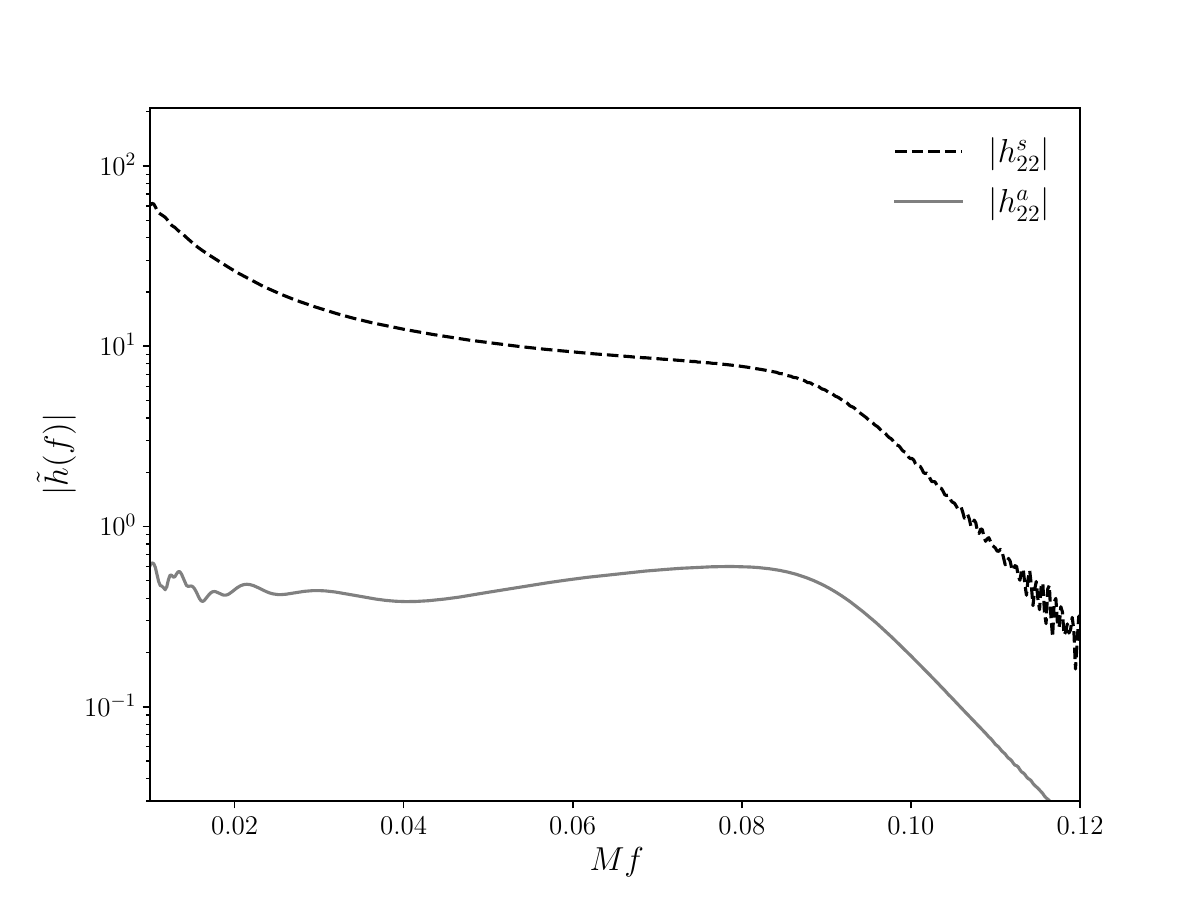}
\caption{The amplitude of the $(l=2,m=2)$ symmetric and anti-symmetric waveform components in the frequency-domain co-precessing frame 
for a $(q=1,\chi_1=0.4,\theta_{LS}=60^{\circ})$ NR simulation.}
\label{fig:NRFDsym_asym}
\end{figure}

The third property we took from the inspiral expressions (\ref{eq:PNsym}) and (\ref{eq:PNasym}) was that the 
anti-symmetric amplitude can be considered as a rescaling of the symmetric amplitude. Fig.~\ref{fig:NRFDsym_asym} 
illustrates that this holds for the entire waveform. It shows the frequency-domain ampiltude of the symmetric
and anti-symmetric (2,2) contributions for a configuration with $(q=1, \chi=0.4, \theta_{\rm LS} = 60^\circ)$,
case \texttt{CF\_7} in Ref.~\cite{Hamilton:2023qkv}.
We see in particular that the turnover to ringdown occurs at the same frequency (the ringdown frequency is 
at $fM \sim 0.09$ for this configuration). We also see that, as discussed above, the symmetric and anti-symmetric
contributions decay at the same rate. 

We will now exploit these features to construct a model for the anti-symmetric amplitude and phase.

\subsection{Structure of the model}

Based on the observations in the previous section we construct an approximate model of the 
anti-symmetric contribution to the (2,2) multipole in the co-precessing-frame as follows. We start with a model for the 
symmetric contribution, which provides us with the symmetric amplitude, $A_s (f)$, and symmetric phase, $\phi_s(f)$.
In the examples we consider here the symmetric contribution is calculated from NR waveforms. In a full model, 
we would start with the symmetric contribution from an already existing model. An explicit example is the 
multi-mode precessing-binary model described in Ref.~\cite{Thompson:2023}, but the anti-symmetric model described 
in this paper can be applied to any existing frequency-domain 
precessing-binary model that 
separately provides $A_s (f)$, and symmetric phase, $\phi_s(f)$, and the precession angle $\alpha(f)$, as we describe 
below.

To construct the anti-symmetric amplitude, $A_a(f)$, we model the ratio $\kappa(f) = A_a(f) / A_s(f)$. In the 
inspiral we use a post-Newtonian estimate of the amplitude ratio, and find that we can model the amplitude
ratio accurately through to the ringdown by adding only one higher-order term, which we fit to our NR data. In the 
ringdown we treat the amplitude ratio as a constant, as motivated in the previous section. The amplitude model is presented
in Sec.~\ref{sec:amplitude}

To construct the anti-symmetric phase, during the inspiral we combine the symmetric phase, $\phi_s(f)$, and the 
precession angle $\alpha(f)$,
as prescribed by Eq.~(\ref{eq:PNasym}), i.e., $\phi_a(f) \sim \phi_s(f)/2 + \alpha(f)$. 
In the merger-ringdown the anti-symmetric phase will behave as $\phi_a(f) \sim \phi_s(f)$. We apply an overall
time and phase shift to smoothly connect these two functional forms at some transition frequency. We note that
we find that it is possible to produce an accurate model of the anti-symmetric phase $\phi_a(f)$ using the same
transition function for any binary, i.e., the parameters of the transition function do not need to be fit across the 
binary parameter space. The phase model is presented in Sec.~\ref{sec:phase}

\section{Numerical relativity data}
\label{section:NR_data}

The model is tuned to 80 single-spin NR waveforms generated using the BAM code~\cite{Bruegmann:2006ulg,Husa:2007hp}. 
They cover mass ratios $q = \{1, 2, 4, 8\}$,
spin magnitudes on the larger black holes of $\chi_1 = \{0.2, 0.4, 0.6, 0.8\}$, and angles of spin misalignment with the orbital 
angular momentum of $\theta_{\rm LS} = \{30^\circ, 60^\circ, 90^\circ, 120^\circ, 150^\circ\}$. In labelling the configurations, the 
cases are ordered according to the 
mass ratios, then the spin magnitudes, then the misalignment angles; for example, case 57 corresponds to $(q=4, \chi_1 = 0.8,
 \theta_{\rm LS} = 60^\circ)$. This is the same indexing as in Ref.~\cite{Hamilton:2023qkv}, which provides full details on the catalog
 of simulations. To motivate the model and test our modelling assumptions we have also used families of simulations that consider
 variations in the initial in-plane-spin direction, based on those in Ref.~\cite{Kalaghatgi:2021inv}. One notable addition to this
 set were simulations of superkick configurations where the black holes were given an additional out-of-plane momentum, to 
 remove a secular drift in the centre-of-mass.
 
Several processing steps are performed to prepare the NR data for modelling, using the tools in Ref.~\cite{London:2015,positive:2020}.
We start with the waveform data for the $\ell=2$ multipoles of the Weyl scalar, $\psi_{4,2m}$, in an inertial frame. We apply a Hann window
to remove ``junk'' radiation from the beginning of the waveforms (a non-physical artefact of the initial data), and to
remove numerical noise in the post-ringdown waveform. Furthermore, to ensure that the frequency-domain step-size is sufficiently small, 
the time-domain data are padded with zeros to the right. More details are given in Ref.~\cite{Hamilton:2023qkv}.

We then transform to a frame that tracks $\hat{J}(t)$; as described in Ref.~\cite{Hamilton:2021pkf}, this retains the approximation that the 
direction of $\hat{J}(t)$ is constant throughout the waveform. Modelling deviations from this approximation are left to future
work, and are discussed further in Ref.~\cite{Hamilton:2021pkf}, along with the procedure to track $\hat{J}(t)$ and perform this transformation. 
At the level of approximation and accuracy of the anti-symmetric model presented here, we do not expect this approximation to have
any appreciable effect on the final model. 
We then transform the $\psi_{4,2\pm2}$ multipoles to a co-precessing frame, the ``quadrupole-aligned'' (QA)~\cite{Schmidt:2010it} or 
``optimal emission direction''~\cite{OShaughnessy:2011pmr,Boyle:2011Gapc} frame.
In this frame the multipoles are significantly simplified, with the $\left( l=2, |m|=2 \right)$ multipoles having the strongest amplitude and the 
precession-induced modulations minimised. 
 
The $(2,\pm2)$ multipoles of time-domain co-precessing-frame $\Psi_4$ are now transformed to the frequency domain, 
\begin{equation}
\tilde{\psi}^{CP}_{4,2\pm2}(f) = \int \psi^{CP}_{4,2\pm2}(t) e^{-2\pi i f t}\, dt . \label{eq:FFT}
\end{equation}
To compute the strain, we note that $\Psi_4(t) = - \ddot{h}(t)$, so in the frequency domain we may write,
\begin{equation}
\tilde{h}^{CP}_{2,\pm2}(f)=-\frac{\tilde{\psi}^{CP}_{4,2\pm2}(f)}{\omega^2}, \label{eq:strain_fd}
\end{equation}
where $\omega=2\pi f$. 
The anti-symmetric and symmetric components of the waveform in the QA frame are computed from
\begin{eqnarray}
\tilde{h}_{s}^{NR}(f)  &=&\frac{1}{2}( \tilde{h} ^{CP}_{2,2}+ \tilde{h}^{*CP}_{2,-2}), \label{eq:sym} \\
\tilde{h}_{a}^{NR}(f) &=& \frac{1}{2}( \tilde{h}^{CP}_{2,2}- \tilde{h}^{*CP}_{2,-2}). \label{eq:asym}
\end{eqnarray}
The symmetric and anti-symmetric strains are complex quantities that can be written as
\begin{eqnarray}
\tilde{h}_{s}^{NR}(f) & = &  A_s^{NR}(f) e^{i \phi^{NR}_s(f)},  \label{eq:sym_nr}\\
\tilde{h}_{a}^{NR}(f) & = & A_a^{NR}(f) e^{i \phi^{NR}_a(f)} 
\end{eqnarray}  
and we can easily calculate their amplitude, $A^{NR}$, and phase, $\phi^{NR}$, as their absolute value and argument, respectively. 
We denote the ratio between the anti-symmetric and symmetric amplitudes as $\kappa_{\rm NR}(f) = A_a^{NR}(f) / A_s^{NR}(f)$.

\section{Model of the amplitude ratiio} 
\label{sec:amplitude}

We model the amplitude of the anti-symmetric contribution $A_a(f)$ as a ratio of the symmetric contribution $A_s(f)$, i.e., 
$A_a(f) = \kappa(f) A_s(f)$. 

Our first step in constructing the ratio model is to compute the ratio in the framework of PN theory as a PN expansion in 
terms of $v/c$, where $v$ is the relative velocity of the two black holes and $c$ is the speed of light, and we choose 
geometric units where $c=1$. 
We again restrict our analysis to single-spin binaries. 

To compute the PN ratio, we follow a procedure similar to the illustrative calculation in Sec.~\ref{sec:background}. 
We obtain from Ref.~\cite{Arun:2008kb} the complex 1.5PN expressions of the ($\ell=2, |m|=2)$ multipoles, $h^{PN}_{2\pm2}$, for spinning, precessing binaries with generic inclination angle $\iota$ moving on nearly circular orbits. 
The strains of the $\ell=|m|=2$ multipoles can then be transformed to a co-precessing frame that follows the instantaneous 
orbital plane. To achieve this, we choose a simple co-precessing frame where we set to zero the inclination angle of the orbital angular momentum relative to the 
total angular momentum, $\iota=0$, and also set the precession angle, $\alpha = 0$. We then use an approximate 
reduction to a single-spin system~\cite{Hamilton:2021pkf},
\begin{eqnarray}
\chi_{s/a,x} &=& -\chi \sin(\theta_{LS} - \iota)\cos(\alpha)/2
\\
\chi_{s/a,y} &=& -\chi \sin(\theta_{LS} - \iota)\sin(\alpha)/2
\\
\chi_{s/a,z} &=& \chi \cos(\theta_{LS} - \iota)/2
\end{eqnarray}
where $\chi_{s}=(\chi_1+\chi_2)/2$, $\chi_{a}=(\chi_1-\chi_2)/2$ are the symmetric and anti-symmetric spins defined in Ref.~\cite{Arun:2008kb} and $\theta_{LS}$ is the inclination of the spin from the orbital angular momentum vector. 

We then find that the symmetric and anti-symmetric amplitudes are, 
\begin{eqnarray}
A_s^{PN}(f) & = & -\frac{4 M}{21 D_L}\sqrt{\frac{\pi}{5}}v^2\eta \left( 42 + 84\pi v^3 + v^2(-107+55\eta)  \right. \nonumber \\
                    & & \left.  \ \ \ \ \ \ \ \ \ \ \ \ \ \ \  - 28v^3(1+\delta -\eta)\chi \cos \theta_{LS} \right), \label{eq:PNsymamp}
\\
A_a^{PN}(f) & = & -\frac{2M}{D_L}\sqrt{\frac{\pi}{5}} v^4 (1+\delta) \eta \chi \sin\theta_{LS}. \label{eq:PNasymamp}
\end{eqnarray} 
The ratio of the two amplitudes is then given by,
\begin{widetext}
\begin{equation}
\label{eq:PNratio}
\kappa_{\rm PN}(f) = \frac{21 v^2 (1 + \delta) \chi \sin\theta_{LS}}{2(42+84 \pi v^3 + v^2 (-107 + 55 \eta) - 28 v^3 (1 + \delta - \eta) \chi \cos \theta_{LS})},
\end{equation}
\end{widetext}
where $\eta=m_1m_2/M^{2}$ is the symmetric mass ratio and $\delta=(m_1-m_2)/M>0$ is a fractional mass difference.  
The expression of the PN ratio of the anti-symmetric amplitude over the symmetric amplitude depends on the symmetric mass ratio, $\eta$, the spin magnitude, $\chi$, and the angle $\theta_{LS}$ of the system. Consequently, Eq.~(\ref{eq:PNratio}) can be used to compute the PN ratio of any configuration as a function of frequency since $v=(\pi f M)^{1/3}$.

We cannot expect the PN estimate of the amplitude ratio, Eq.~(\ref{eq:PNratio}), to be accurate through merger and ringdown. 
To capture this behaviour we investigated the addition of unknown higher-order terms and fit their coefficients to the NR data. 
The simplest approach is to add only one additional term, for example, 

\begin{equation}
    \kappa(f)=\kappa_{\rm PN}(f) \, (1+bv^n),
    \label{eq:kappa}
\end{equation}
where $b$ is fit to the NR data. To do this it was necessary to choose an appropriate frequency range 
over which to perform the fit. The NR data are noisy in the early inspiral and in the post-ringdown frequencies, so we first identified
a consistent definition of a frequency range that could be used for all 80 NR simulations, based on the ringdown frequency, 
$f_{RD}$ of each NR configuration. The frequency range that we used in the final fits was given by 
$M f_{min}=\left(M f_{RD}-0.01\right)/5$ and $M f_{max} = M f_{RD}-0.002$. 

\begin{figure}[ht!]
	\includegraphics[width=\linewidth]{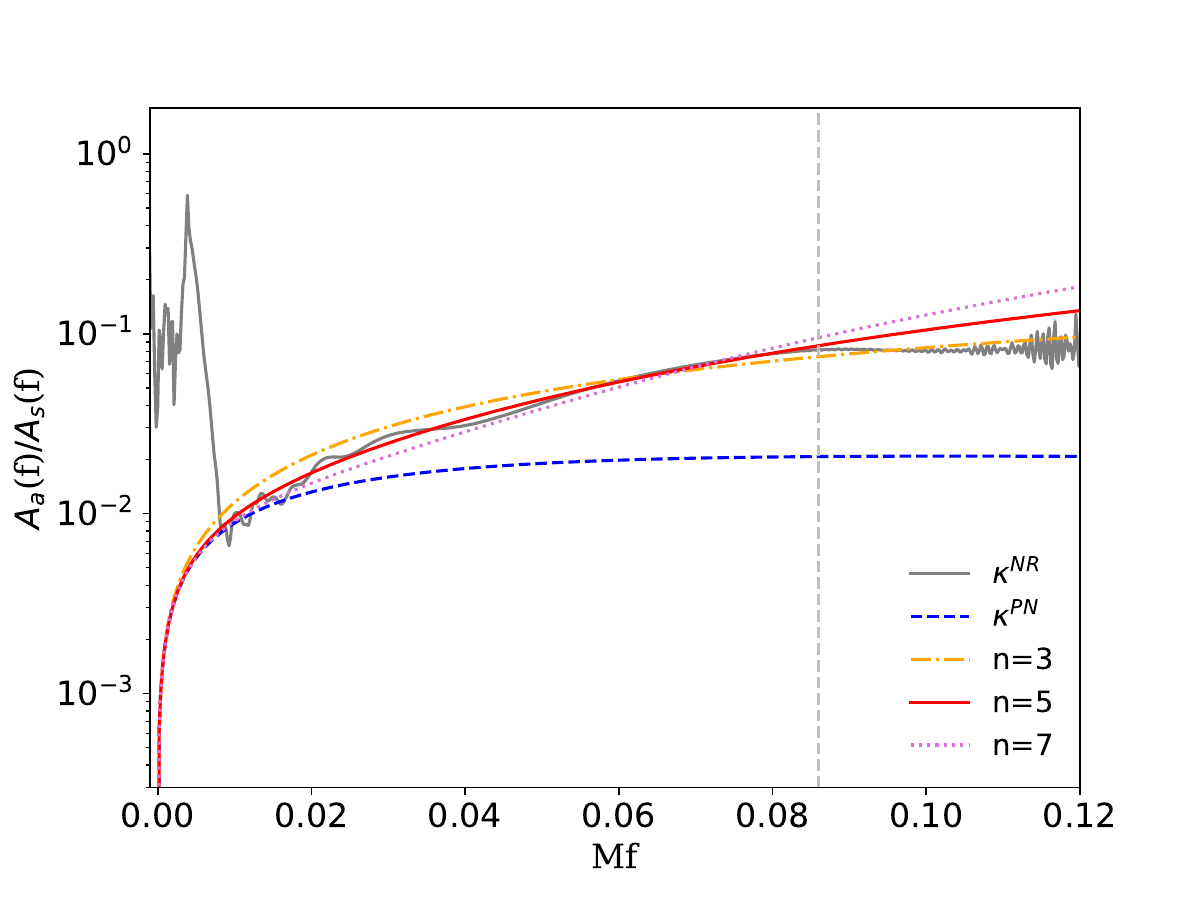}
	\caption{Amplitude ratio for the $(q=1,\chi=0.4,\theta_{LS}=60^o)$ configuration,
	for the NR data, the PN ratio $\kappa_{\rm PN}$, and fits to higher-order corrections as in Eq.~(\ref{eq:kappa}),
	with $n = 3, 5, 7$. The vertical dashed line indicates the ringdown frequency.}	  
	\label{fig:PNcorrections}
\end{figure}

We investigated fits to the amplitude ratio of the form Eq.~(\ref{eq:kappa}) with $n = 3, 5, 7$. Fig.~\ref{fig:PNcorrections} shows the
results for one configuration, and illustrates that $n = 5$ provides the most accurate fit. In fact, we find that $n=5$ consistently provides the 
most accurate fit across all 80 NR simulations. 

It is also clear from Fig.~\ref{fig:PNcorrections} that our fit is not accurate beyond the ringdown frequency. Beyond this point
the amplitude ratio appears to plateau. This is consistent with our expectations from perturbation theory, as discussed in
Sec.~\ref{sec:imr}, which predicts that the amplitude ratio will be constant throughout ringdown. 
To include this feature in our model, we fix $\kappa(f)$ to
the value $\kappa(f_{RD})$ at frequencies $f > f_{RD}$. To avoid a sharp transition we use a moving average algorithm such that,
\begin{equation}
\kappa_n(f) = \frac{1}{(2k+1)}\sum_{i=n-k+1}^{n+k+1} \kappa(f_i), \;\; n \in [k,N-k].
\end{equation}
We use a symmetric window of an equal number of points ($k = 40$) on either side of the frequency $f$ to calculate the moving average. 
Here $N$ is the length of the frequency series and the algorithm updates $\kappa(f)$ for $40 \leq n \leq N-40$.

\begin{figure}[ht!]
\includegraphics[width=\linewidth]{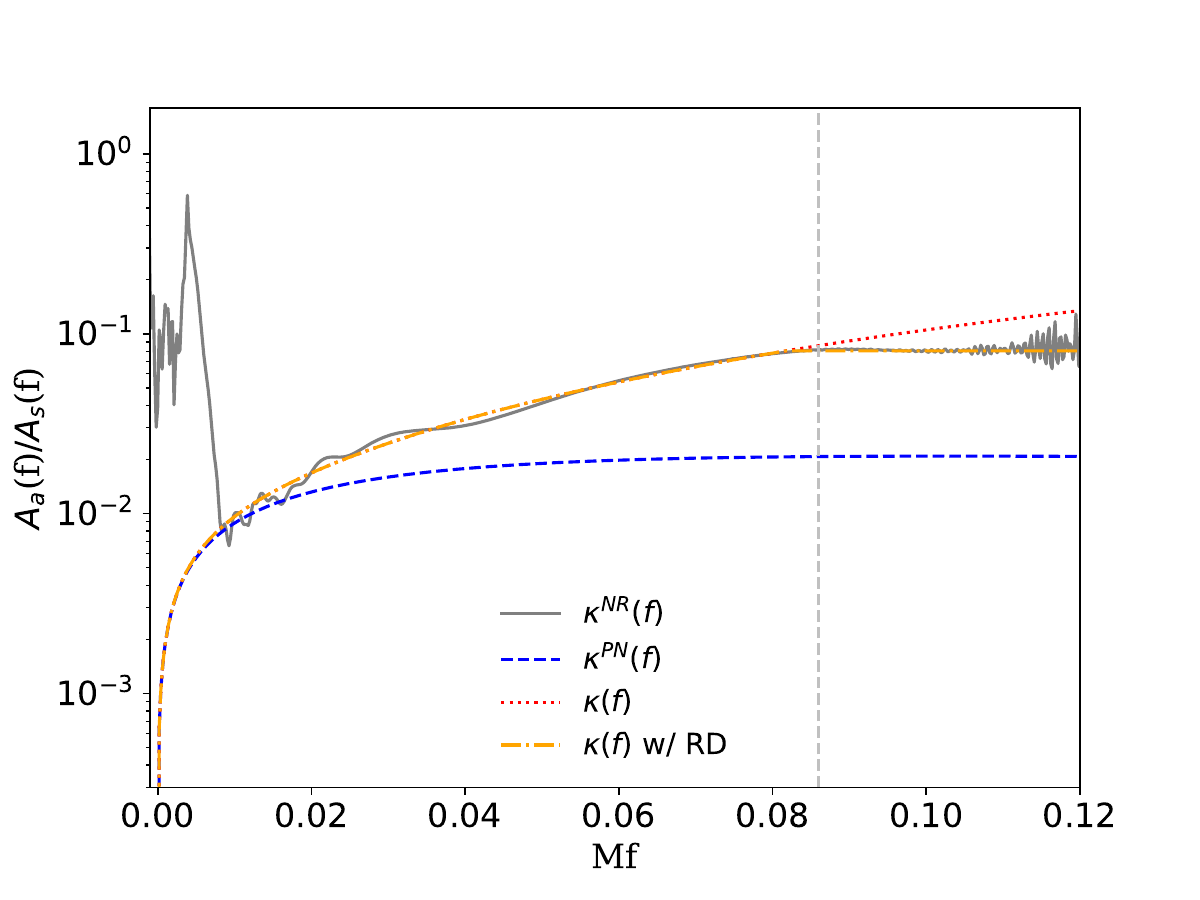}
\includegraphics[width=\linewidth]{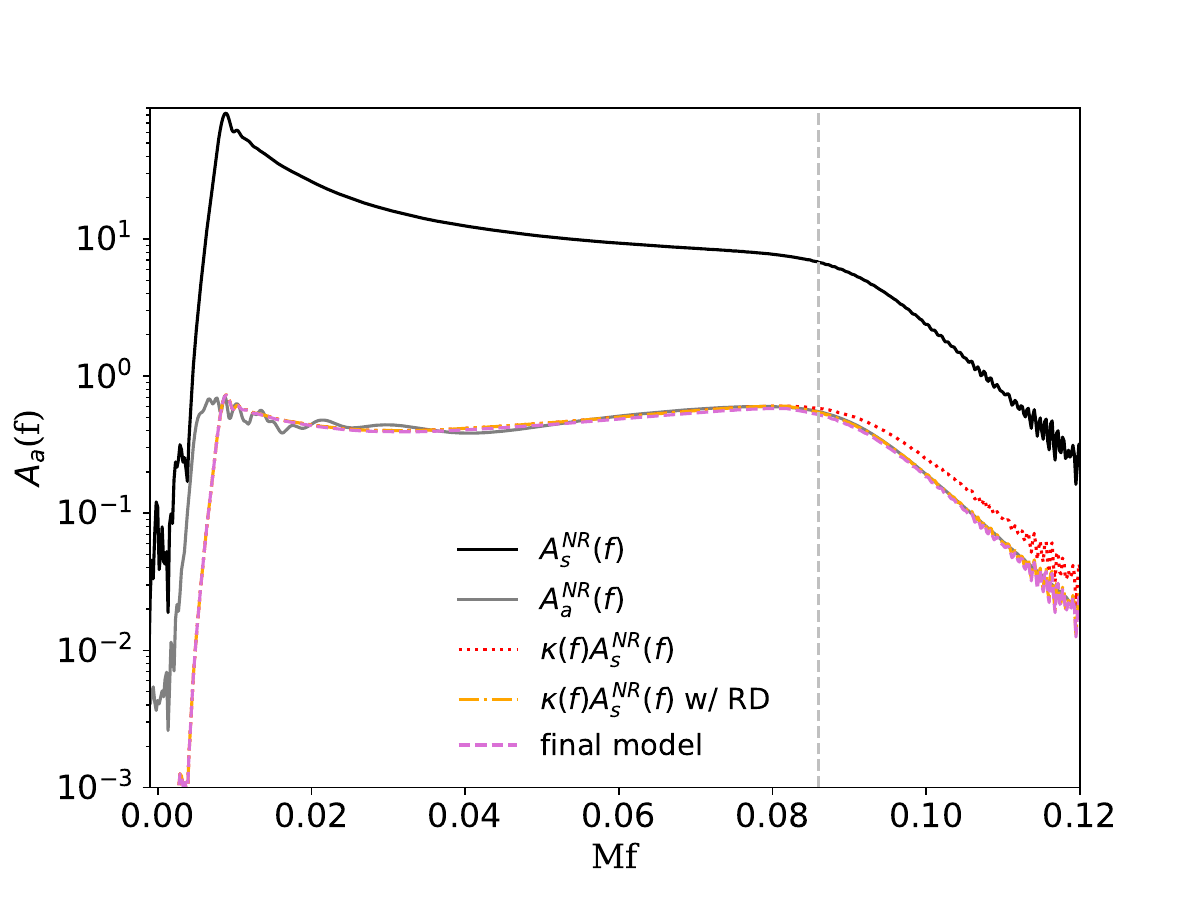}
\caption{The amplitude ratio (top) and amplitude (bottom) for the $(q=1,\chi=0.4,\theta_{LS}=60^\circ)$ configuration. The vertical line indicates
the ringdown frequency for the $(\ell=2,|m|=2)$ multipoles.
}	  
\label{fig:ratiofit1}
\end{figure}

We fit the coefficient $b$ in Eq.~(\ref{eq:kappa}) (with $n =5$) to each of the $80$ single-spin NR waveforms from the NR catalogue in Ref.~\cite{Hamilton:2023qkv}. Fig.~\ref{fig:ratiofit1} shows the resulting fit for the amplitude ratio for the $(q=1,\chi=0.4,\theta_{LS}=60^{\circ})$ configuration. We see
that the final fit, including the levelling off of the amplitude ratio through ringdown, agrees well with the NR data. The lower panel of the figure
shows the resulting estimate for the anti-symmetric amplitude. We see that this agrees well with the NR data up to the point where the NR amplitude becomes
dominated by noise. 

\begin{figure}[ht!]
\includegraphics[width=\linewidth]{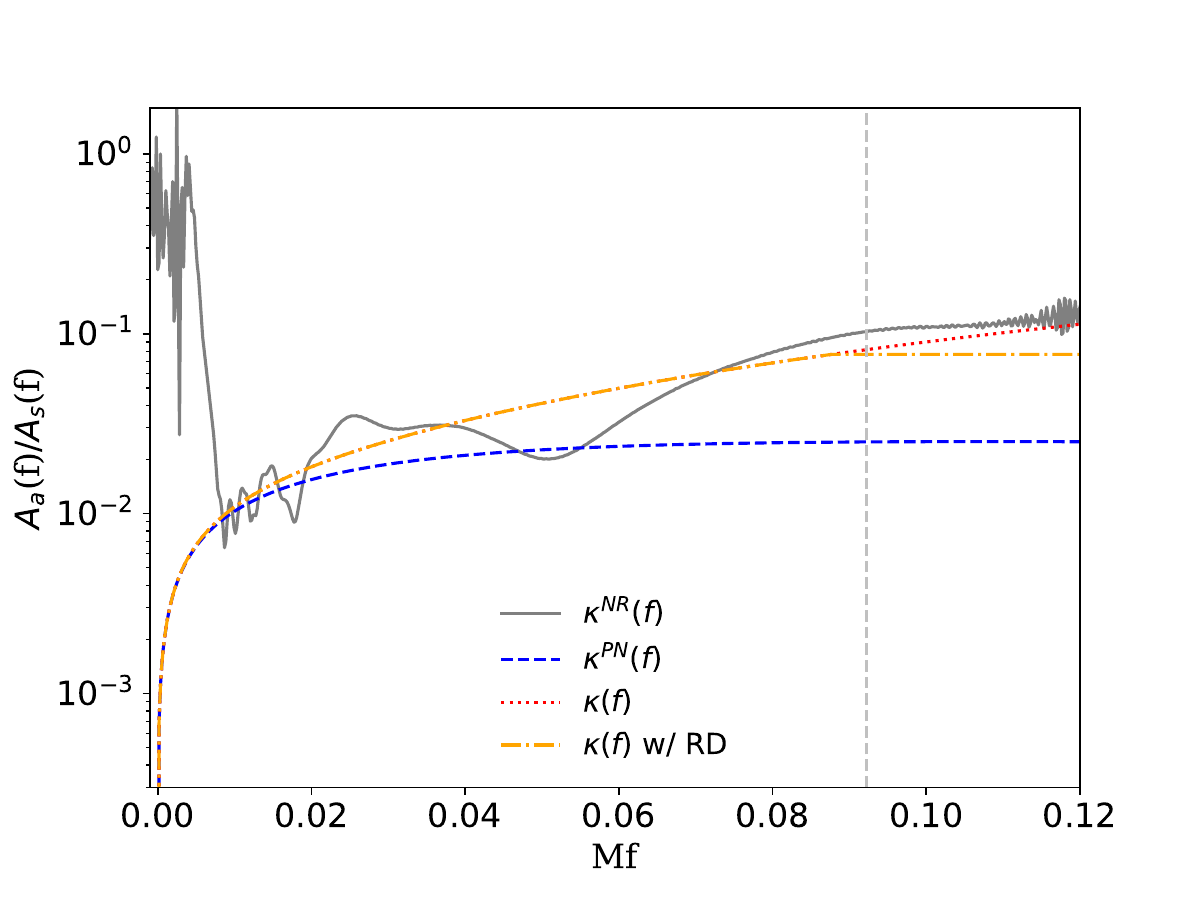}
\includegraphics[width=\linewidth]{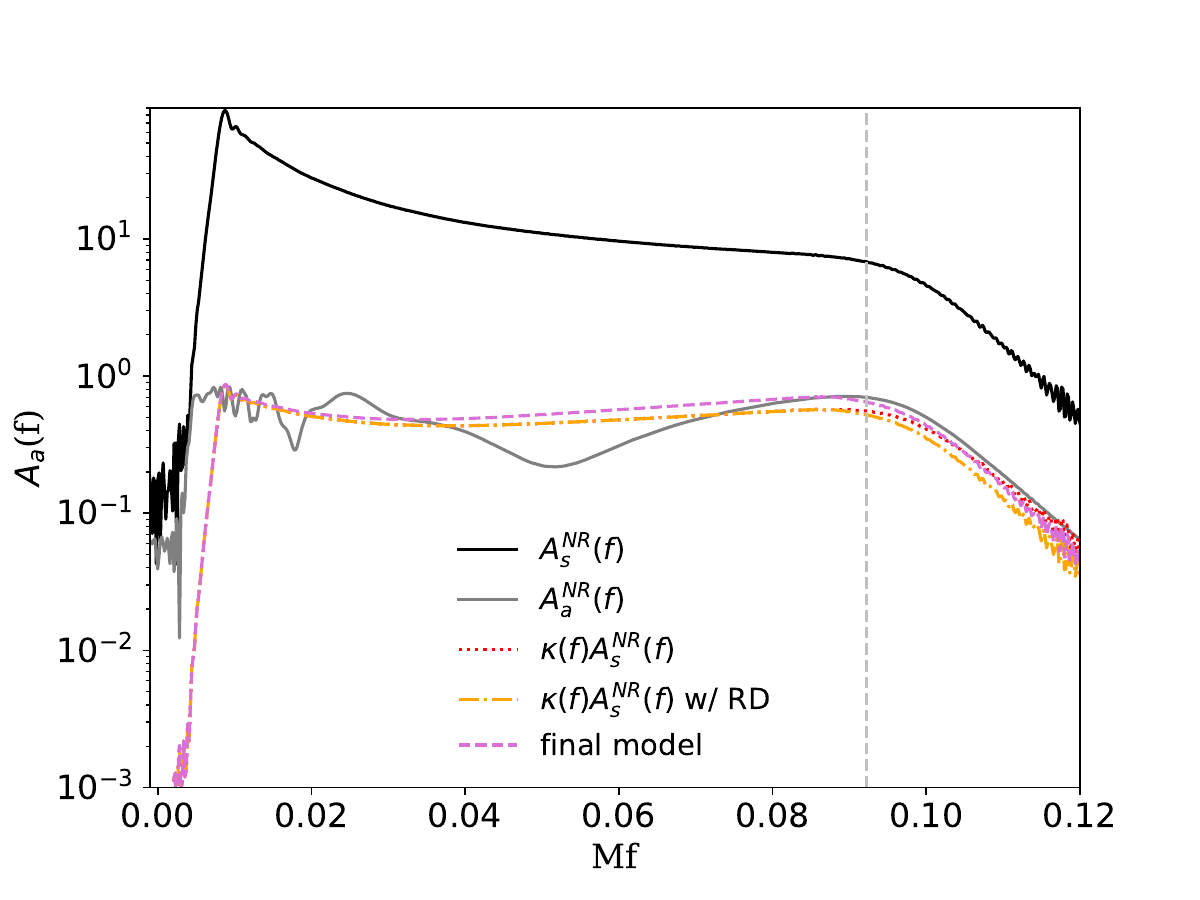}
\caption{ The amplitude ratio (top) and amplitude (bottom) for the $(q=1,\chi=0.8,\theta_{LS}=30^{\circ})$ configuration. The vertical line indicates
the ringdown frequency for the $(\ell=2,|m|=2)$ multipoles.
}	  
\label{fig:ratiofit2}
\end{figure}

The fit cannot reproduce the NR amplitude ratio in all cases. In many cases the NR amplitude ratio oscillates during the inspiral. 
Fig.~\ref{fig:ratiofit2} shows an extreme example of this feature, from the $(q=1,\chi=0.8,\theta_{LS}=30^{\circ})$  configuration. 
It is not clear what causes these
oscillations. Oscillations in the co-precessing-frame amplitude ratio during the inspiral can be due to the choice of 
co-precessing frame, as we will discuss later. However, if that were the cause of the oscillations in the NR data, we would expect 
there to be some correlation with the 
degree of precession in the configuration. Instead we do not find any relationship between the amplitude of the oscillations, which in some cases
are negligible (as in Fig.~\ref{fig:ratiofit1}), and in others (like Fig.~\ref{fig:ratiofit2}) lead to significant variations in the amplitude ratio. We do find, though, that
our model passes through a reasonable estimate of the mean of the oscillations. The largest impact is on the constant value of the amplitude
that the model settles on for the ringdown regime; in the example in Fig.~\ref{fig:ratiofit2} the model's estimate of the anti-symmetric amplitude during the ringdown is roughly 20\% below the NR value. 

\begin{figure}[ht!]
\includegraphics[width=\linewidth]{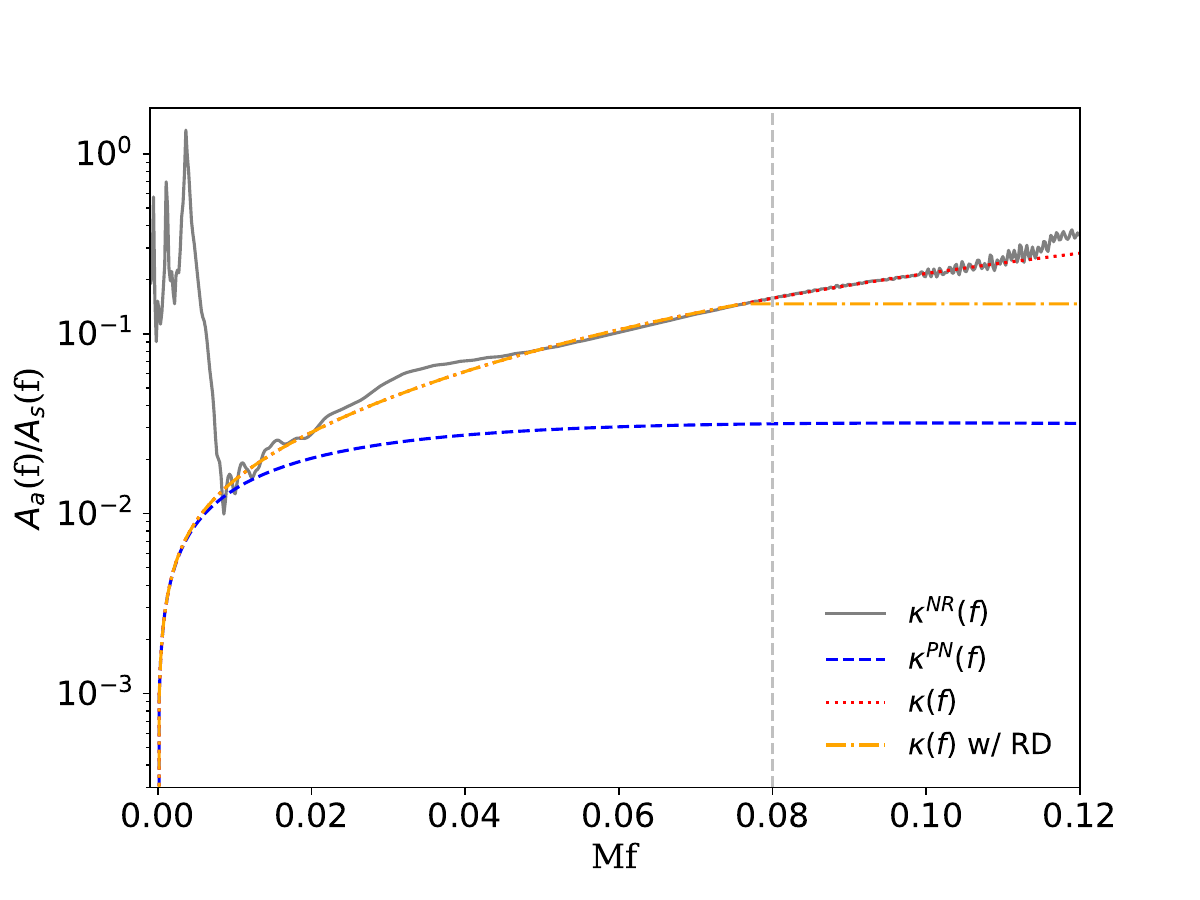}
\includegraphics[width=\linewidth]{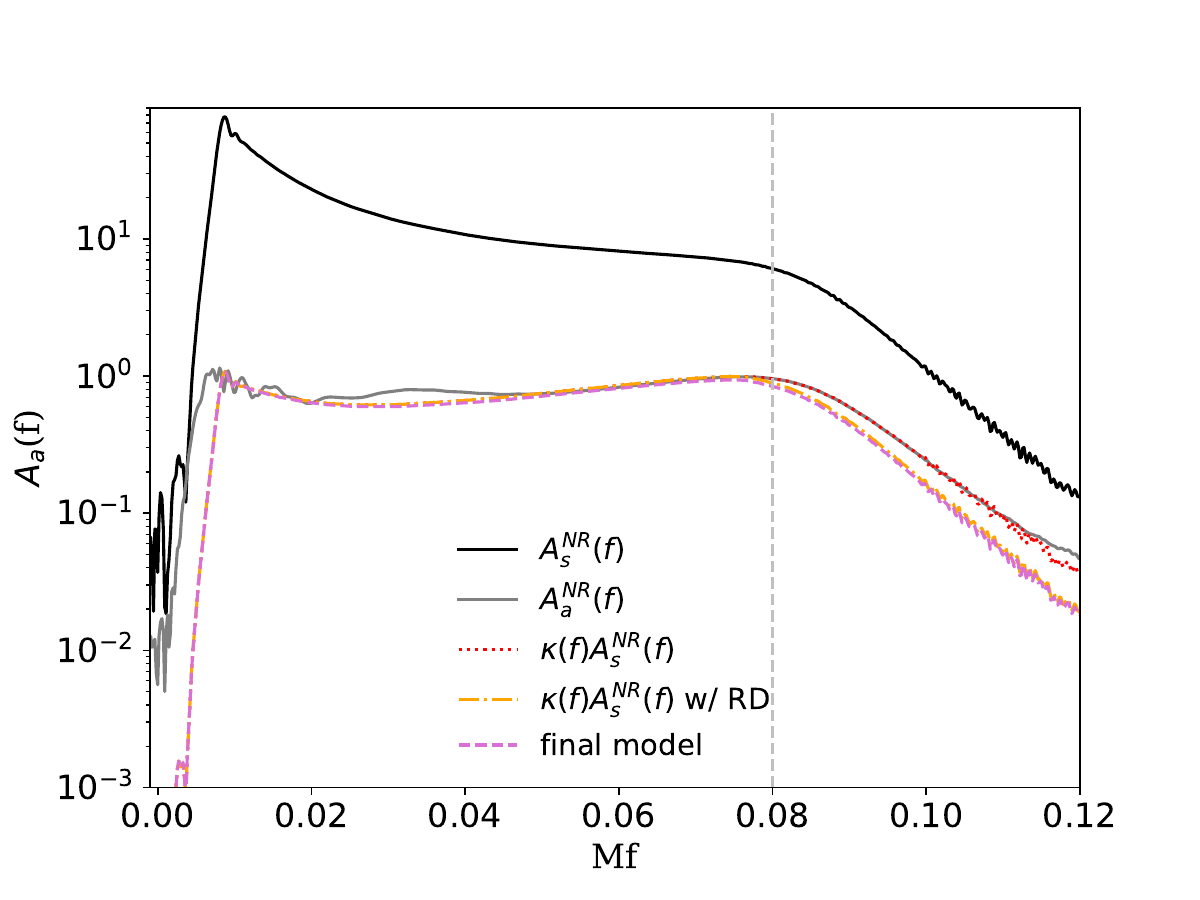}
\caption{
The amplitude ratio (top) and amplitude (bottom) for the $(q=2,\chi=0.4,\theta_{LS}=90^{\circ})$ configuration. The vertical line indicates
the ringdown frequency for the $(\ell=2,|m|=2)$ multipoles.
}	  
\label{fig:ratiofit3}
\end{figure}

Finally, in some cases we also found that the amplitude ratio in the NR data did not level off during the ringdown. Fig.~\ref{fig:ratiofit3} is
an example of this. We have not been able to determine the reason for this. 
As noted previously, we expect that since the $(2,2)$ and $(2,-2)$ multipoles decay at the same rate, that the ratio between their amplitudes
would remain constant during the ringdown. It is possible that this effect is obscured by numerical noise. Regardless of the cause, 
and in the absence of any compelling explanation for alternative behaviour, we have chosen to impose the behaviour that we expect from 
perturbation theory.

\begin{figure}[ht!]
    \centering\includegraphics[width=\linewidth]{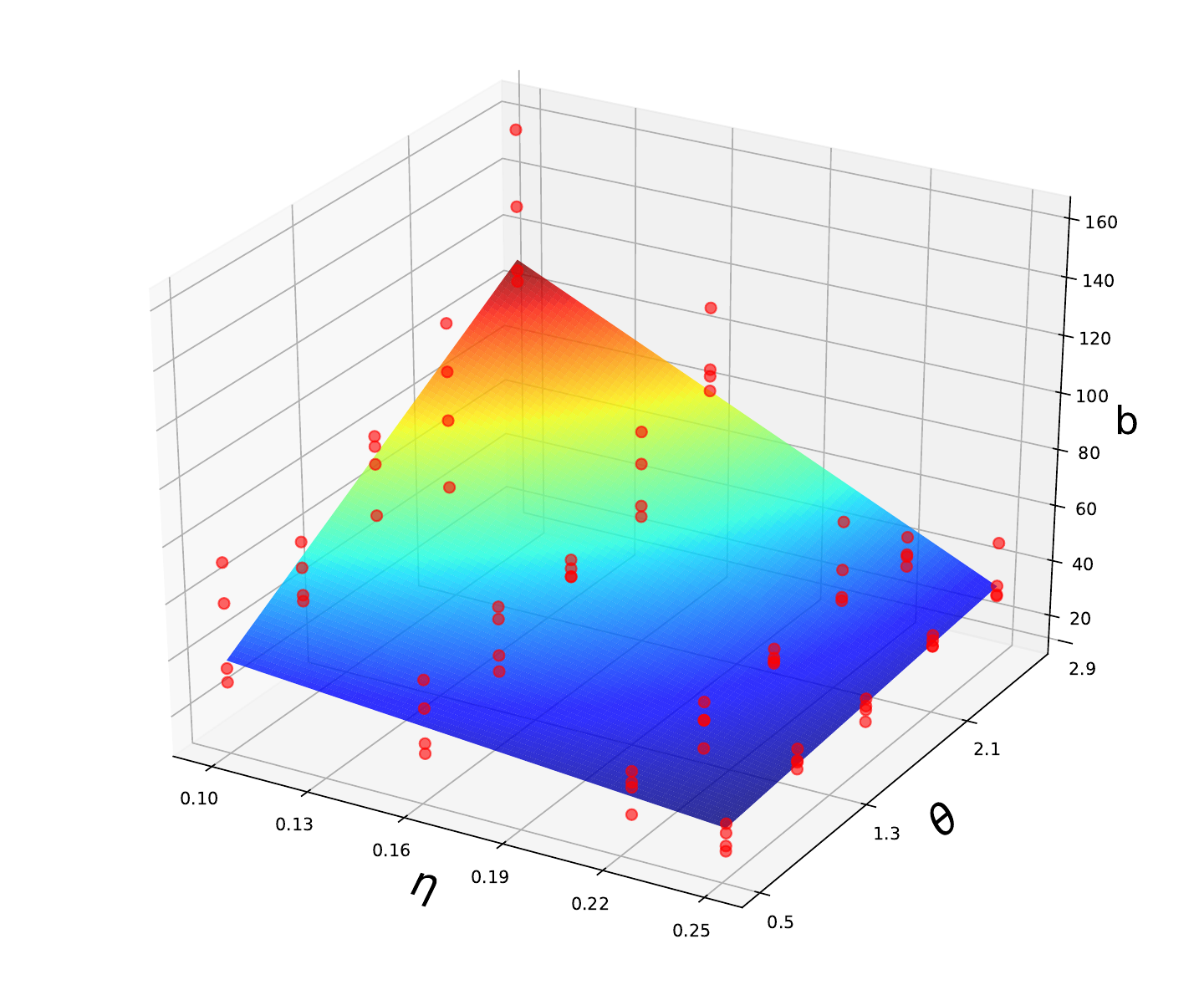}
\caption{Surface $b(\eta, \theta_{LS}) = b_0 +  b_1 \eta + b_2 \theta_{LS} + b_3 \eta \theta_{LS}$ fit of the model's coefficient, $b$, to the two-dimensional parameter space ${\eta,\theta_{LS}}$. The red points denote the $80$ computed $b$ coefficients of the multipole asymmetry amplitude model.  
}	  
\label{fig:surfacefit}
\end{figure}

\begin{figure}[ht!]
    \centering\includegraphics[width=\linewidth]{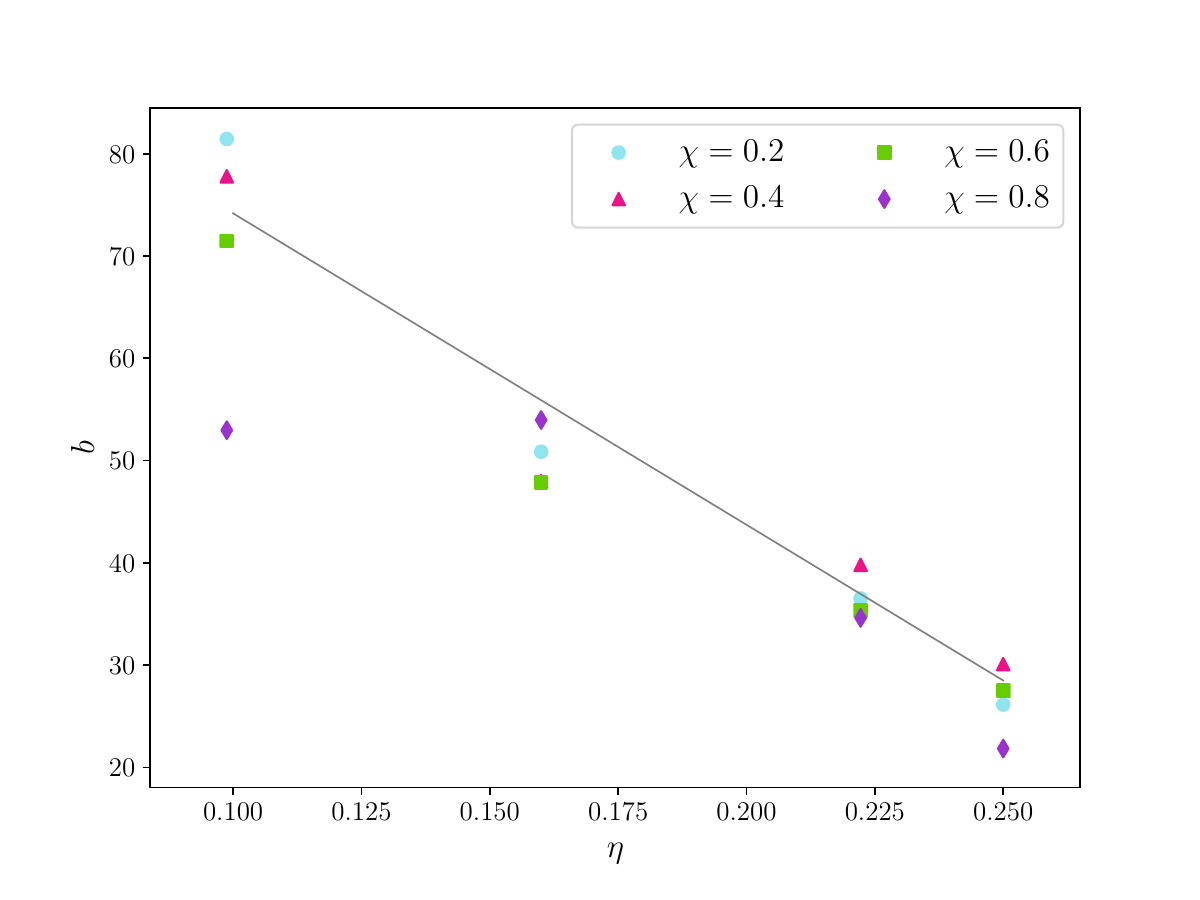}
\caption{The $b$ coefficient as a function of the symmetric mass ratio, $\eta$, for a selected angle $\theta_{LS}=90^o$ and all the available spin values, $\chi=[0.2,0.4,0.6,0.8]$. The grey line shows the surface fit, $b(\eta,90^\circ)$ from Eq.~(\ref{eq:bcoef}). 
}	  
\label{fig:indepchi}
\end{figure}

The coefficient $b$ is shown as a function of symmetric mass ratio and misalignment angle in Fig.~\ref{fig:surfacefit}; the values for all four spin
magnitudes are shown together. This plot illustrates the general trend of $b$ across the parameter space. We find that there
is no general trend with respect to the spin magnitude, and this is illustrated more clearly in Fig.~\ref{fig:indepchi}, which shows
$b$ as a function of symmetric mass ratio, for each spin magnitude, for the subset of cases with $\theta_{\rm LS} = 90^\circ$.  The PN amplitude
ratio already includes a linear dependence on the spin magnitude, and given the uncertainty in our fits, we do not attempt to 
identify a higher-order spin dependence. We motivate this point further in Sec.~\ref{sec:matches} (Fig.~\ref{fig:matches_surface}). We then include all simulations into a two-dimensional parameter-space fit 
of the form,
\begin{equation}
b(\eta, \theta_{\rm LS}) = b_0 +  b_1 \eta + b_2 \theta_{LS} + b_3 \eta \theta_{LS}
\label{eq:bcoef}
\end{equation} 
where we find $b_0 = 18.0387$, $b_1 = 15.4509$, $b_2 = 55.1140$ and $b_3 = -203.6290$.
From Eq.~(\ref{eq:bcoef}), we notice that $b$ does not go to zero when $\theta_{LS}$ is $0^\circ$ or $180^\circ$. 
However, the presence of the $\sin \theta_{LS}$ term at the numerator of the ansatz of the ratio model in Eq.~(\ref{eq:PNratio})
ensures that the multipole asymmetry goes to zero at these points. The amplitude as predicted by this fit is shown on each of our
figures and labelled as ``final model''. 

Two caveats to this approach are worth noting. One is the choice of co-precessing frame. Previous work has shown that the 
symmetric (2,2) contribution takes a simple form in the QA co-precessing frame; indeed, the amplitude evolution can be approximated
by the (monotonic) amplitude of an equivalent aligned-spin configuration. This is not necessarily the case for the anti-symmetric 
contribution, and this is one possible cause of the oscillations that we see (although, as noted previously, it does not show a clear
correlation with the strength of precession). Conversely, we found that the amplitude evolution of 
the anti-symmetric (2,2) multipole was monotonic for PN waveforms if we choose $\iota = \alpha = 0$ in their construction
(which is equivalent to choosing a co-precessing frame that tracks the Newtonian orbital angular momentum, i.e., the normal
to the orbital plane), but if we consider the PN waveforms in the QA frame then the anti-symmetric (2,2) amplitude shows strong
modulations. This illustrates that the anti-symmetric component can depend strongly on the choice of co-precessing frame, and
although we do not expect this to be a significant issue at the level of accuracy or approximation of the current model, it may 
require better understanding in future refinements of anti-symmetric models. 

The second point is that our model is constructed based on the phenomenology of single-spin binaries. If the model is to be
used for generic binaries, one much choose a method to treat two-spin configurations. One option, which is employed in
the \texttt{PhenomX} implementation~\cite{Hamilton:2021pkf,Thompson:2023}, is as follows. 

We can describe the spin of an equivalent single-spin system using Eqs.~(16) and (17) in Ref.~\cite{Hamilton:2021pkf}, 
but diverge slightly in the definition of the in-plane spin as follows:
\begin{equation}
\chi^\perp =
\begin{cases}
\chi_{\rm{as}}\cos^2(\theta_q) + \chi_{\rm{p}}\sin^2(\theta_q), & 1\leq q \leq 1.5\\
\chi_{\rm{p}}, & q>1.5
\end{cases}
\end{equation}
where $\chi_{\rm{p}}$ is the effective precession spin as defined in Ref.~\cite{Schmidt:2015tmgII}. The anti-symmetric amplitude for 
an equal-mass binary with both spins equal in magnitude, entirely in-plane and pointing in the same direction, must drop to zero. 
The symmetric in-plane spin combination of Eq.~(15) in Ref.~\cite{Hamilton:2021pkf} cannot account for this effect in superkick
configuration. Therefore, we can instead use an anti-symmetric in-plane spin combination,
\begin{equation}
\chi_{\rm{as}} = \frac{|\bf{S}^\perp_1-\bf{S}^\perp_2|}{m_1^2},
\end{equation}
to appropriately map two-spin to single-spin systems for generating the anti-symmetric waveform.

%

\section{Phase model} 
\label{sec:phase}

\begin{figure}[ht]
\includegraphics[width=\linewidth]{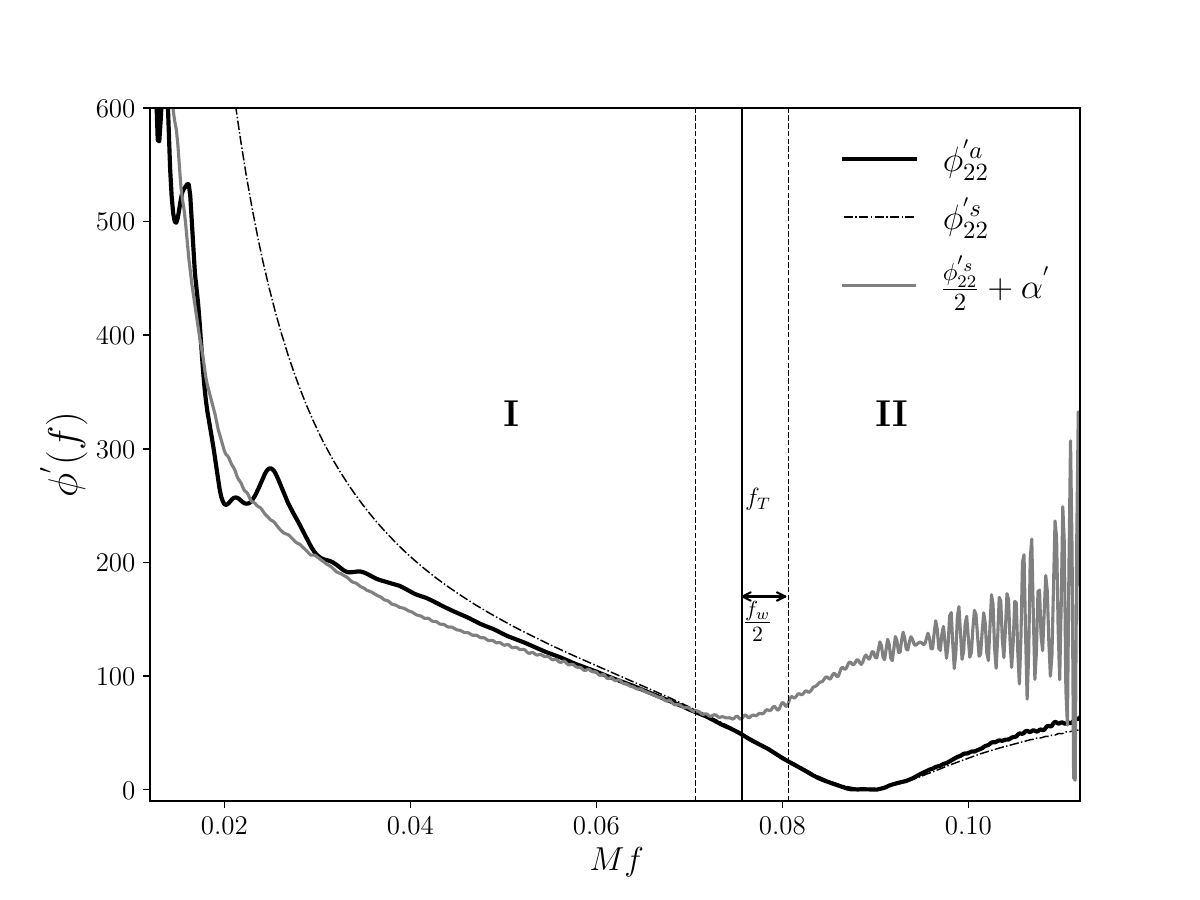}
\caption{The anti-symmetric phase derivative (thick black line), the symmetric phase derivative (grey dashed line) and the 
combination of half of the symmetric phase derivative with the derivative of the precession angle $\alpha$ (thin grey line). The 
specific configuration shown here is for a single-spin binary with $(q=1$, $\chi = 0.2$, $\theta_{LS}=30^{\circ})$.}  
\label{fig:asymphd}
\end{figure}

GW phases for chirp signals in the frequency domain are typically quite featureless and therefore not conducive to modeling. 
Following a suite of models for the symmetric phase~\cite{Husa:2015iqa,Khan:2015jqa,Pratten:2020fqn}, we focus first on
the anti-symmetric phase derivative. Fig.~\ref{fig:asymphd} demonstrates that the anti-symmetric phase derivative 
(i.e., frequency) behaviour in the time-domain as discussed in Sec.~\ref{sec:imr} is preserved in the frequency domain as well. Remarkably, we find that it is possible to construct a map of the symmetric phase derivative to the anti-symmetric phase derivative 
that is independent of the binary's parameters. Therefore, we do not need to produce any parametric fits for the map over 
the intrinsic parameter space of BH binaries, which makes this model extremely simple.

Our model of the anti-symmetric phase derivative is defined by the piecewise function, 
\begin{equation}
\phi_a'(f) =
\begin{cases}
\frac{1}{2} \phi'_s(f) + \alpha'(f) + A_0, & f \leq f_T -\frac{f_w}{2} \\
\phi'_{\rm{int}}(f), &  f_T -\frac{f_w}{2} < f \leq  f_T +\frac{f_w}{2} \\
\phi'_s(f), &  f_T+\frac{f_w}{2} \leq f < 0.12,
\end{cases}
\end{equation} 
where the phase derivative in the intermediate region is given by,
\begin{equation}
\begin{split}
\phi'_{\rm int}(f) =  \frac{1}{2} \bigg[1-\frac{3}{2f_w}((f-f_T) - \frac{(f-f_T)^3}{3f_w^2})\bigg] \big(\frac{\phi'_s}{2}  + \alpha ' + A_0\big) \\
     +\frac{1}{2} \bigg[1+\frac{3}{2f_w}((f-f_T) - \frac{(f-f_T)^3}{3f_w^2})\bigg] (\phi'_s + B_0)
\end{split}
\label{eq:phiinter}
\end{equation}
As is evident from Fig.~\ref{fig:asymphd}, the functional form of region I transitions to the functional form of region II in the 
intermediate region. To ensure smooth transition in the intermediate region an obvious choice would be a $tanh$ windowing 
function. Noting that a $tanh$ function can be computationally inefficient during model evaluation, we instead use the Taylor 
expansion of the $tanh$ function up to second order with appropriate normalization to construct the phase derivative functional 
form of Eq.~(\ref{eq:phiinter}). 

The phase derivative ansatz was calibrated to NR simulations by treating the transition frequency, $f_T$, the width of the 
transition window, $f_w$,  and the phase coefficients ($A_0$ and $B_0$) as free parameters. The window parameters were 
not particularly sensitive to the tuning and the minor variations could be attributed to the noise in the anti-symmetric phase 
derivatives; $f_T = 0.85 f_{RD}$ and $f_w = 0.005$ was found to be an optimal choice across the entire single-spin parameter space. 

Once the parameters of the window function were fixed, we investigated the impact of fixing the phase coefficients. A best 
fit to data for $B_0=0$ was found to be consistent across the parameter space; $A_0$ on the other hand showed some variation, 
but no specific trend. Furthermore, choosing the algebraic mean of $A_0$ for the set of $80$ simulations did not significantly 
impact the quality of the fit. This shows (1) the fitting algorithm tries to find the best $A_0$ for continuity of the phase derivative 
at $f_T$, and (2) the variation in optimal $A_0$ across the parameter space is more likely due to noisy data and not a function 
of intrinsic parameters of the binary.

\begin{figure}[ht]
\includegraphics[width=\linewidth]{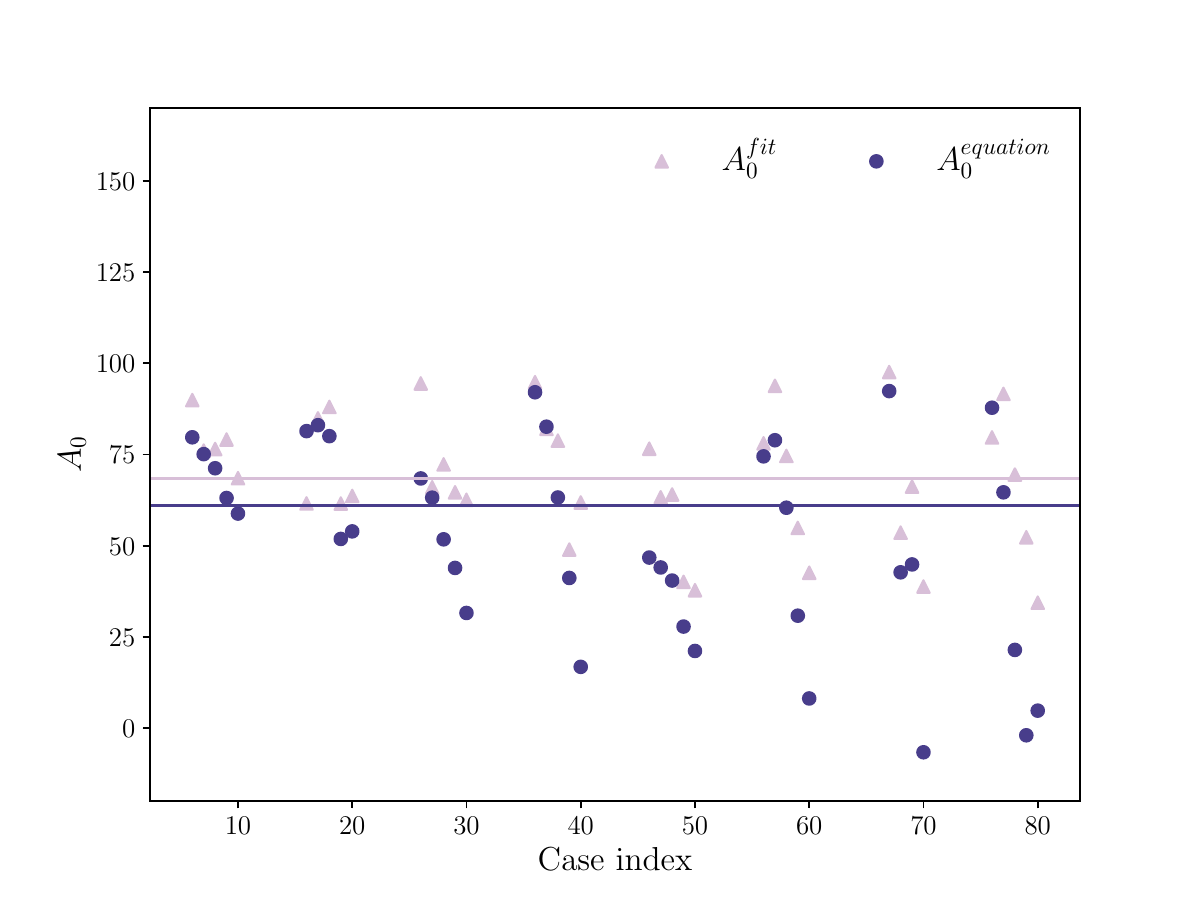}
\caption{Comparison of phase coefficient $A_0$, obtained from fitting the phase ansatz to NR data (light purple triangles) 
across the parameter space and from NR data using Eq.~(\ref{eq:a0}) (blue circles), for cases with $\chi = 0.4$ and 0.8. 
The algebraic mean of the set of coefficients are shown by horizontal lines in corresponding colors. The case index is as described in Sec.~\ref{section:NR_data}.}
\label{fig:a0fit}
\end{figure}

Applying a shift to the phase derivative is equivalent to an overall time shift of the waveform. We exploit this freedom by fixing 
the symmetric phase derivative minima to be $0$ at $f_{RD}$. This imposes,
\begin{equation}
A_0 = \frac{1}{2} \phi'_s(f_T) - \alpha'(f_T) .
\label{eq:a0}
\end{equation}
Figure~\ref{fig:a0fit} shows that  $A_0$ obtained from the fitting algorithm and from NR data using Eq.~(\ref{eq:a0}) reasonably well for a majority of the cases (cf. \textit{y-axis} on Fig.~\ref{fig:asymphd}). The anti-symmetric model will be used with a phenomenological symmetric waveform model, so an $A_0$ derived from 
the symmetric waveform model makes the phase construction self-consistent and robust. Furthermore, Fig.~\ref{fig:a0fit} 
highlights that the gain in accuracy by making a model to capture the near-stochastic behaviour of $A_0$ may be 
outweighed by errors 
introduced by over-modelling. As such, our model of the anti-symmetric phase is NR-informed but, noting that the data for the 
anti-symmetric waveform is often close to numerical noise, we prioritised our understanding of the physics rather than 
model optimization.

The phase of the anti-symmetric waveform is obtained by integrating the two pieces,
\begin{equation}
\phi_a(f) =
\begin{cases}
\frac{1}{2} \phi_s(f) + \alpha(f) + A_0f + \phi_{A0}, & f \leq f_T \\
\phi_s(f) +  \phi_{B0}, &  f_T\leq f < 0.12,
\end{cases}
\label{eq:phiafinal}
\end{equation} 
where the integration constant $\phi_{B0}$ is determined by continuity at $f_T$
\begin{equation}
\phi_{B0} = \alpha(f_T) - \phi_s(f_T) + A_0f_T.
\label{eq:phiB0}
\end{equation}
Finally, the phase of the asymmetry is modulated by the in-plane spin direction $\alpha$; therefore, the initial phase, $\phi_{A0}$, is the value of $\alpha$ at a reference frequency i.e., $\phi_{A0} = \alpha(f_{\rm{ref}})$.

\section{Model accuracy}
\label{sec:matches}

A standard measure of waveform accuracy used extensively in the literature is the \textit{match} of the waveform model with NR waveforms, defined as,
\begin{equation}
M(h_{NR},h_M) = 4 \, \rm{Re}\int \limits_0^\infty \frac{h_{NR}(f)h^*_M(f) }{S_n(f)}df , \label{eq:matchdefn}
\end{equation}
where $h(f) = h_+(f) - i h_\times(f)$ is a complex frequency sequence constructed from the two polarizations of the waveform. 
As such, calculating matches of just the anti-symmetric waveform is not physically meaningful. Furthermore, a true measure 
of performance of precessing waveforms in data analysis can only be obtained by calculating matches of the full waveform 
in the inertial frame, making considerations for precession as well as extrinsic parameters. Therefore, matches of just the 
anti-symmetric waveform in the co-precessing frame provide some indication of the accuracy of one ingredient in the 
full waveform, but do not indicate the overall accuracy of the corresponding precessing waveform. 

\begin{figure*}[ht]
\includegraphics[width=\textwidth]{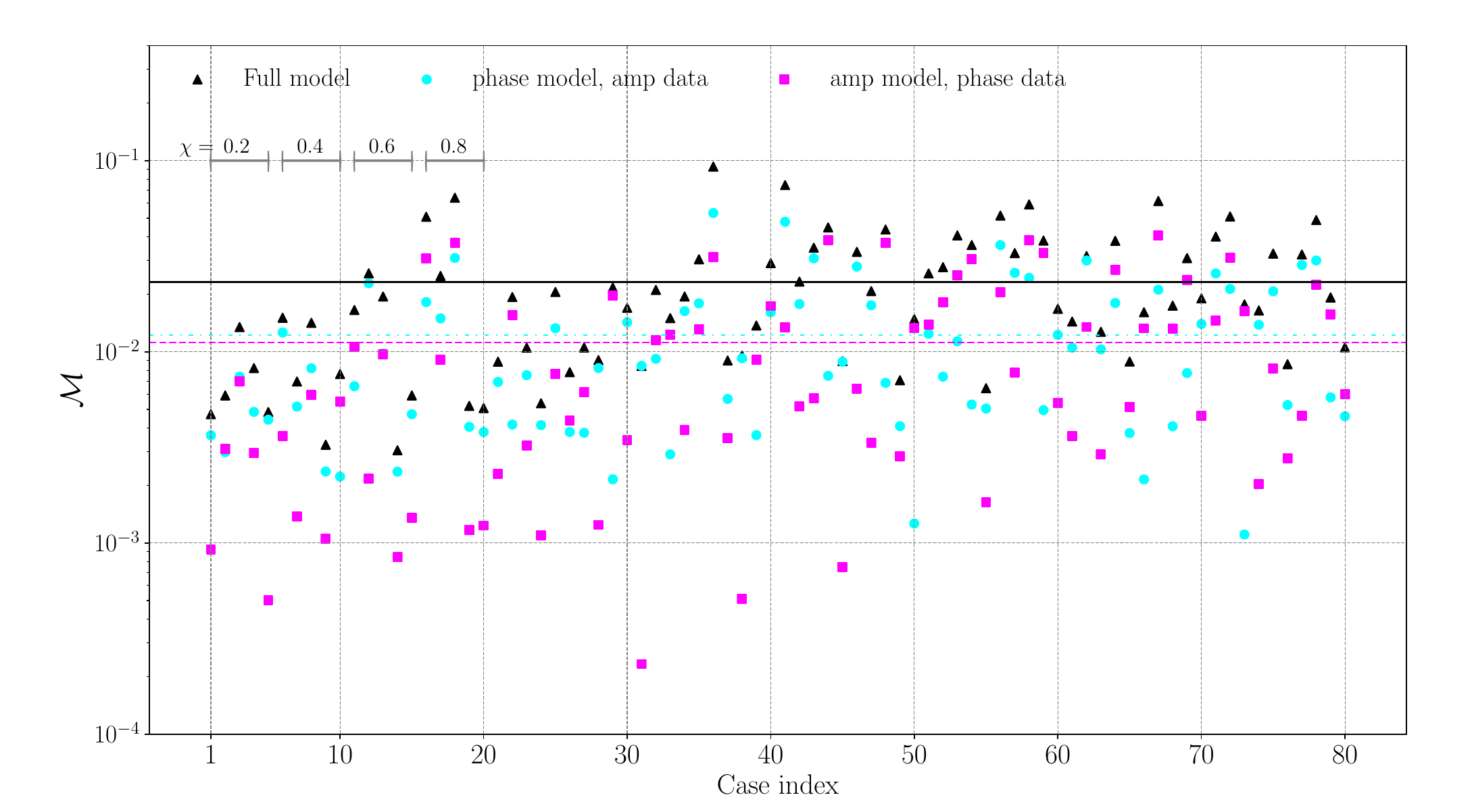}
\caption{\textit{Mismatches} of the anti-symmetric waveform model in the co-precessing frame with NR data. The black triangles show the mismatches for the combined amplitude and phase model while the magenta squares (cyan circles) show the mismatches for just the amplitude (phase) model with the phase (amplitude) constructed from NR data. The dashed magenta and cyan lines show the average mismatch for the amplitude and phase model, respectively; on average they are of comparable accuracy. The black solid line shows the average mismatch for the overall model. The \textit{x-axis} denotes the case index of the NR simulation as usual i.e., five different $\theta_{\rm{LS}}$ for each spin magnitude shown in figure, for $q=1$,$2$,$4$ and $8$.}
\label{fig:matches}
\end{figure*}

However, an inner product like that in Eq.~(\ref{eq:matchdefn}) is a useful measure of agreement between two complex frequency 
series. Since a match-like calculation for the anti-symmetric contribution in the co-precessing frame cannot be interpreted in terms
of either signal detection efficiency or parameter measurement accuracy, there is no reason to include the detector sensitivity,
and so we will use a simpler inner product of the form,
\begin{equation}
\langle h_{NR} | h_M \rangle =  \rm{Re} \int \limits_{f_1}^{f_2} h_{NR}(f)h^*_M(f) df , \label{eq:innerproduct}
\end{equation} where $f_1$ is the starting frequency of the NR waveform in geometric units,
 $M f_1 = 0.02$, and 
$M f_2 = 0.15$, after which point the amplitude of the NR waveform is below the noise floor of the data. 
We consider normalised waveforms, $\hat{h} = h / \sqrt{\langle h | h \rangle}$, so that the maximum value of the inner
product is one. We used the standard implementation of this inner product in  \texttt{pycbc}~\cite{alex_nitz_2021_4849433}, a python software package for GW data analysis, for our match computations. We then consider the ``mismatch'' between the two waveforms, 
\begin{equation}
\mathcal{M} = 1 - \frac{\langle h_{NR} | h_M \rangle}{\sqrt{ \langle h_{NR} | h_{NR} \rangle
\langle h_M | h_M \rangle}}.
\end{equation} 

In Fig.~\ref{fig:matches} we show the mismatches of the anti-symmetric waveform constructed from our model with the 80 
NR waveforms that were used to calibrate the model. To determine the accuracy of the individual components, we also 
computed matches of the amplitude (phase) model complemented by phase (amplitude) constructed from NR data, 
with the full anti-symmetric waveform constructed from NR data. The overall accuracy of both the models are comparable. We note that although the model was
verified using the same waveforms as used for modelling, since the NR tuning was relatively simple --- i.e., a single co-efficient
in Eq.~(\ref{eq:kappa}) fit to the four-parameter ansatz Eq.~(\ref{eq:bcoef}) across a two-dimensional parameter space --- 
using a much smaller subset of waveforms would have produced a model of similar accuracy, and the simplicity of this
model and the single-spin parameter space obviates any concerns about over-fitting or unexplored corners of parameter space.

\begin{figure*}[ht]
\includegraphics[width=\textwidth]{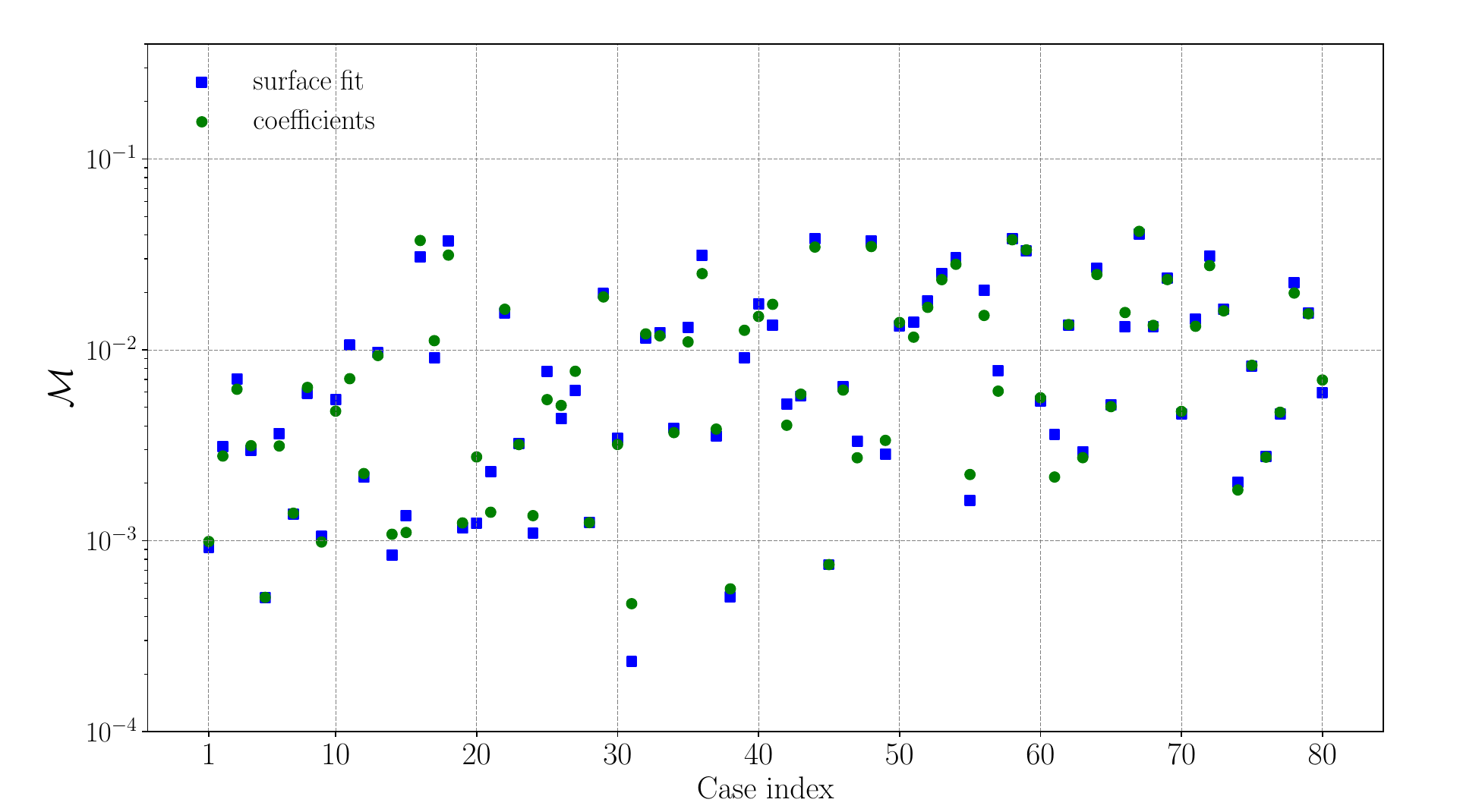}
\caption{\textit{Mismatches} (same as Fig.~\ref{fig:matches}) showing comparison of the amplitude model constructed using the spin magnitude-independent surface fit of Eq.~(\ref{eq:kappa}) (\textit{blue squares}) with the amplitude model constructed from the true fit coefficients (\textit{green circles}).}
\label{fig:matches_surface}
\end{figure*}

To investigate the quality of the surface fit in Eq.~(\ref{eq:bcoef}), we also computed mismatches for the amplitude model using fit 
coefficients $b$ of Eq.~(\ref{eq:kappa}). As is evident from Fig.~\ref{fig:matches_surface}, for most cases the performance is unchanged and for the handful of cases where the mismatch changes, the difference is not very significant. This further illustrates that capturing the non-linear dependence on spin magnitude is unlikely to make significant improvement to the amplitude model. In addition to the argument made for using Eq.~(\ref{eq:a0}) to 
calculate $A_0$ from the symmetric waveform and precession angle, $\alpha$, we calculated mismatches for the different choices of 
$A_0$ in the phase model -- i.e., $A^{fit}_0 $and $A^{equation}_0$ in Fig.~\ref{fig:a0fit} -- to confirm that there was no reduction 
in model accuracy. 

Note that the anti-symmetric waveform model is downstream from the symmetric waveform model as  well as the precession angle models. Therefore, enhancement in performance of the overall model due to the addition of an asymmetry model must always be discussed in the context of the underlying symmetric, precessing waveform model. This is beyond the scope of present work and will be discussed in the context of the current generation frequency-domain precessing-binary \texttt{PhenomXPNR} model ~\cite{Thompson:2023}.

\section{Conclusion}
\label{section:conclusion}

We have presented a method to model the anti-symmetric contribution to the $(\ell=2, |m|=2)$ multipoles in the co-precessing frame. 
This is a general approach that can be applied to any frequency-domain model that provides the symmetric contribution to the (2,2) 
amplitude, phase and precession angle, $\alpha$. We expect that the main insights of this model can be used to easily generalise
the procedure to the time-domain and be useful in including multipole asymmetry in current generation time-domain models~\cite{Estelles:2021gvs,Ramos-Buades:2023ehm}.
We summarise the key insights here. 

For the amplitude, we observe that the anti-symmetric amplitude can be easily modeled as a rescaling of the symmetric amplitude. As shown in
Fig.~\ref{fig:NRFDsym_asym}, both amplitudes have the same basic structure, in particular the same ringdown frequency and decay
rate. In the inspiral our model of the amplitude ratio is based on a PN expression for single-spin systems, and in the late inspiral and merger
a higher-order correction to the PN expression is calibrated to 80 NR simulations of single-spin binaries~\cite{Hamilton:2023qkv}. In
the ringdown we make use of the prediction from perturbation theory that the ratio of the symmetric and anti-symmetric amplitudes will 
be constant. The amplitude model ratio is presented in Eqs.~(\ref{eq:PNratio}), (\ref{eq:kappa}) and the fit to the higher-order contribution 
given in Eq.~(\ref{eq:bcoef}). 

For the phase, we use the facts that in the inspiral the frequency of the anti-symmetric contribution equals the orbital frequency plus the 
spin precession frequency, and that during the ringdown the symmetric and anti-symmetric frequencies are the same. We are able to construct
a mapping from the symmetric phase and precession angle, $\phi_s$ and $\alpha$, to the anti-symmetric phase, $\phi_a$, motivated
by the 80 single-spin simulations, but without the need of any explicit tuning. The model of the phase is given in Eqs.~(\ref{eq:a0}), (\ref{eq:phiafinal}) and 
(\ref{eq:phiB0}). An additional crucial observation is that a rotation of the initial in-plane spin direction of $\Delta \alpha$ introduces a 
corresponding shift of $\Delta \alpha$ into the anti-symmetric phase. It is this observation that allows us to model the dependence on
in-plane-spin direction, even though we do not have a set of NR simulations that span this subspace. 

The procedure presented here is not in itself a signal model. As already noted, one must also provide the symmetric amplitude, phase and
precession angle. In addition, having constructed a model for the anti-symmetric contribution to the (2,2) multipoles in the co-precessing
frame, one must then ``twist them up'' to produce the multipoles in the inertial frame. This has been done for the recent extension of the 
multi-mode frequency-domain precessing-binary \texttt{IMRPhenomXPNR} model~\cite{Thompson:2023}.

This is the first phenomenological full inspiral-merger-ringdown model of the anti-symmetric multipole contributions, and there are many directions for improvement
and issues to be resolved. The first limitation of the model is that it is based only on single-spin binaries. We expect that generic two-spin 
systems can in most cases be modelled to a good approximation by single-spin systems, but given that the anti-symmetric contribution 
is itself a weak effect, it is unclear how well this approximation can be used. It would be useful to study the applicability of the single-spin
approximation for the anti-symmetric contribution, and, indeed, to extend the model to two-spin systems. The anti-symmetric model is also
limited to the $(\ell=2,|m|=2)$ multipoles. We argue in Fig.~\ref{fig:higher_asym} that the (2,2) contribution will be sufficient for most
applications, since the anti-symmetric amplitude is far weaker than the symmetric, but for high signal-to-noise-ratio (SNR) systems, higher-order
anti-symmetric multipoles will ultimately be required. 

When decomposing the signal into a co-precessing frame and corresponding time- or frequency-dependent angles, one will find (depending
on the co-precessing frame chosen) oscillations in the amplitude and/or phase of the anti-symmetric contribution; recall that the 
co-precessing-frame maps exactly to a non-precessing aligned-spin waveform only for the leading-order quadrupole 
terms~\cite{Schmidt:2012tmg,Hamilton:2021pkf,Thompson:2023}. In this work we have removed oscillations in the inspiral amplitude by
simply setting the inclination angle $\iota$ and precession angle $\alpha$ to zero in the PN expressions we have used for the 
signal multipoles~\cite{Arun:2008kb}; oscillations remain in the co-precessing-frame signals for the NR waveforms, but our model
consists of only a monotonic fit through these oscillations. A better understanding of these oscillations, or at least a model that
captures them, is a necessary next step in modelling the multipole asymmetries.

We have assumed that the only affect of the initial in-plane spin direction is to introduce an overall offset in the anti-symmetric phase. This 
approximation will not be exact during the merger-ringdown. In particular, in the ringdown we have made the approximation that the amplitude 
ratio will be independent of the initial in-plane spin direction. However, the amplitude ratio will be determined by the relative phase of the 
symmetric and anti-symmetric contributions when ringdown begins, and we have not attempted to model this effect. 

These limitations aside, we find that our method to construct the anti-symmetric contribution has an average mismatch error better than 0.03. 
Given that the anti-symmetric contribution is in general less than 10\% of the total SNR, we expect that its contribution to the total mismatch
uncertainty of a model will be less than $3\times10^{-3}$, which is below the average mismatch error of current Phenom and SEOBNR
precessing-binary models~\cite{Ramos-Buades:2023ehm,Thompson:2023}. With the addition of this first anti-symmetric model, we expect
that current models will be able to make more precise measurements of black-hole spins and gravitational recoil. The accuracy of
these measurements will depend significantly on the underlying symmetric model, as well as where a signal lies in the binary parameter
space, and we leave such accuracy studies with individual models to future work. 

 \section{Acknowledgements}
We thank Jonathan Thompson, Eleanor Hamilton and Lionel London for many thought-provoking discussions on generic precessing binaries. We are indebted to Lionel London for sharing his numerical relativity data utility package with us, this was used to extract the anti-symmetric waveform data in the co-precessing frame. We thank Frank Ohme and Jannik Mielke for insightful discussions on multipole asymmetry, and Chinmay Kalaghatgi for discussions on in-plane spin rotation in superkick configurations.

S.G thanks Sebastian Khan for stimulating discussions on frequency-domain waveform modeling, Duncan Macleod for excellent inputs on computing practices and Edward Fauchon-Jones for support in using supercomputing facilities.

The authors were supported in part by Science and Technology Facilities Council (STFC) grant ST/V00154X/1 and European Research Council (ERC)
Consolidator Grant 647839. S.G acknowledges support from the Max Planck Society’s Independent Research Group program. P.K was also supported by the GW consolidated grant: STFC grant ST/V005677/1.

Simulations used in this work were performed on the DiRAC@Durham facility, managed by the Institute for Computational Cosmology 
on behalf of the STFC DiRAC HPC Facility (www.dirac.ac.uk). The equipment was funded by BEIS capital funding via STFC capital grants 
ST/P002293/1 and ST/R002371/1, Durham University and STFC operations grant ST/R000832/1. In addition, several of the simulations used in this 
work were performed as part of an allocation graciously provided by Oracle to explore the use of our code on the Oracle Cloud Infrastructure. 

This research also used the supercomputing facilities at Cardiff University operated by Advanced Research Computing at Cardiff (ARCCA) on 
behalf of the Cardiff Supercomputing Facility and the HPC Wales and Supercomputing Wales (SCW) projects. We acknowledge the support 
of the latter, which is part-funded by the European Regional Development Fund (ERDF) via the Welsh Government. In part the computational 
resources at Cardiff University were also supported by STFC grant ST/I006285/1.

Various plots and analyses in this paper were made using Python software packages \texttt{LALSuite}~\cite{lalsuite}, \texttt{PyCBC}~\cite{alex_nitz_2021_4849433}, \texttt{gwsurrogate}~\cite{Field:2013cfa}, \texttt{Matplotlib}~\cite{Hunter:2007}, \texttt{Numpy}~\cite{Harris_2020}, \texttt{lmfit} and \texttt{Scipy}~\cite{2020SciPy-NMeth}.

\bibliography{references.bib}

\end{document}